\documentstyle[feynmf,12pt,titlepage]{article}
\begin{document}

\newtheorem{lem}{Lemma}
\def\e{{\mathrm{e}}}
\def\mathrm{\rm}

\def\beq{\begin{equation}}
\def\eeq{\end{equation}}
\def\bea{\begin{eqnarray}}
\def\eea{\end{eqnarray}}
\def\nn{\nonumber}
\def\dim{{3-2\varepsilon}}
\def\ve{\varepsilon}
\def\meas#1{{d^\dim#1\over(2\pi)^\dim}}
\def\In{{\cal I}}
\def\I{{\mathrm{i}}}
\def\J{{\cal J}}
\def\K{{\cal K}}
\def\M{{\cal M}}
\def\Li2{\mathop{\mathrm{Li_2}}}
\def\Ei{\mathop{\mathrm{Ei}}}

\begin{titlepage}
\begin{flushright}
HU-TFT 96-22\\
hep-ph/9606216\\
June 4, 1996
\end{flushright}
\begin{centering}
\vfill
{\Large\bf Feynman diagrams to three loops in three dimensional field theory}
\vspace{1cm}

Arttu K. Rajantie\footnote{E-mail: Arttu.Rajantie@Helsinki.Fi}
\vspace{0.5cm}

{\em Department of Physics \\ P.O. Box 9 \\ 
FIN-00014 University of Helsinki, Finland}
\vspace{2cm}

{\bf Abstract}

\vspace{0.5cm}
\end{centering}

\noindent
The two point integrals contributing to the self energy of a particle in a
three dimensional quantum field theory are calculated to two loop order in
perturbation theory as well as the vacuum
ones contributing to the effective potential to three loop order. 
For almost every integral an expression in terms of elementary and dilogarithm
functions is obtained. For two integrals, the master integral and the Mercedes
integral, a one dimensional integral representation is obtained with
an integrand consisting only of elementary functions. The results are
applied to a scalar $\lambda\phi^4$ theory.

\vfill
\end{titlepage}

\section{Introduction}
When physical phenomena are described by quantum field theories, all the
observable quantities are expressed in terms of functional integrals.
Since these integrals can be evaluated exactly only in very special
cases, one has to use lattice simulations or analytical approximation methods.
The most common method is perturbation theory.
One expands the
desired quantity as a series of integrals 
represented by Feynman diagrams.

Although the field theories of particle physics are four dimensional,
the importance of three dimensional theories has grown recently.
The main reason is the method of dimensional reduction of a 
four dimensional finite temperature field theory to a three dimensional
zero temperature effective theory 
%\cite{ref:Ginsparg,ref:JacTem,ref:Appelquist,ref:Nadkarni,%
%ref:Landsman,ref:dimred}. 
[1--6].
This technique has been applied to
the electroweak phase transition of the early universe
%\cite{ref:electroweak,ref:lattice,ref:uusi}. 
[7--9].
Three dimensional field
theories are also important in the theory of critical phenomena.

The nature of the phase transition is an important question in
both main applications of three dimensional field theories. Thus 
the effective potential which gives the true ground state of the system has
an essential significance. Unfortunately, in dimensionally reduced effective
theory, perturbative calculations are applicable neither in the 
symmetric phase nor in the immediate 
vicinity of the phase transition. 
However, the perturbative results obtained deep in the broken phase 
can give new insight into the problem and the lattice results 
\cite{ref:lattice,ref:uusi}.

At one loop level the perturbative calculations are fairly easy.
When one needs higher corrections, the integrals get more complicated.
In four dimensions these integrals have been studied recently by many
authors
%\cite{ref:series,ref:series2,ref:Mellin,ref:Gegenbauer,%
%ref:Scharf,ref:Bauberger,ref:Kreimer,ref:Ghinculov,%
%ref:Avdeev,ref:Broadhurst}, 
[10--19]
but a three dimensional discussion has been missing.

The purpose of this paper is to evaluate all the integrals needed for
the self energy of a particle in a three dimensional scalar theory to
two loop order and the ones needed for the effective potential to
three loop order. 
As shown by Weiglein {\it et~al.~}\cite{ref:Weiglein}, 
the self energy of a particle in a gauge field theory can be expressed
in terms of these scalar integrals.
Most of the integrals are calculated in a straightforward
way, but with two integrals a different route must be chosen. 

The paper is organized as follows. 
In Sect. \ref{scalarsect} the simpler integrals are
calculated explicitly. Sect. \ref{mastersect} is devoted to
the two more intricate integrals.
In Sect. \ref{scalarquant} the integrals are applied to the calculation
of the self energy and the effective potential of 
scalar $\lambda\phi^4$ theory.

\section{Evaluation of scalar integrals}
\label{scalarsect}
\subsection{Classification of integrals}
In order to calculate the self energy of a specific system one needs to 
consider all possible one particle irreducible two point diagrams. 
Let us assume that the Lagrangian consists only of terms at most quartic
in the fields. Then the diagrams can be composed from three and four leg
vertices using Feynman rules.
\begin{fmffile}{topo1l}
\begin{figure}
\centering{
$\alpha)$
\fmfframe(0,0)(0,0){
\begin{fmfgraph}(30,30)
\fmfleft{v1}
\fmfright{v2}
\fmf{plain,left}{v1,v2,v1}
\end{fmfgraph}
}
\hskip 1cm
$\beta)$
\scriptsize{
\fmfframe(0,0)(0,0){
\begin{fmfgraph*}(60,30)
\fmfleft{v1}
\fmfright{v2}
\fmf{plain}{v1,i1}
\fmf{plain}{i2,v2}
\fmf{plain,left,label=1}{i1,i2}
\fmf{plain,right,label=2}{i1,i2}
\fmfforce{(.25w,.5h)}{i1}
\fmfforce{(.75w,.5h)}{i2}
\end{fmfgraph*}
}}
\hskip 1cm
\normalsize{$\gamma)$}
\scriptsize{
\fmfframe(0,0)(0,0){
\begin{fmfgraph*}(40,40)
\fmfleft{v1}
\fmfright{v2}
\fmf{plain,left,label=1}{v1,v2}
\fmf{plain,label=2}{v1,v2}
\fmf{plain,right,label=3}{v1,v2}
\end{fmfgraph*}
}}}

\caption{\label{topo1l}Topologies of one and two loop vacuum diagrams and
one loop two point diagram}
\end{figure}
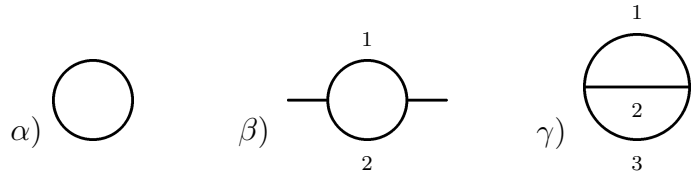

\end{fmffile}

\begin{figure}
\centering{
\begin{tabular}{rr}
\large{a)}
\scriptsize{
\fmfframe(0,0)(0,-30){
\begin{fmfgraph*}(120,120)
\fmfleft{v1}
\fmfright{v2}
\fmf{plain}{v1,i1}
\fmf{plain}{i3,v2}
\fmf{plain,left=0.5,label=1}{i1,i2}
\fmf{plain,left=0.5,label=4}{i2,i3}
\fmf{plain,right=0.5,label=2}{i1,i4}
\fmf{plain,right=0.5,label=5}{i4,i3}
\fmf{plain,label=3}{i2,i4}
\fmfforce{(.5w,.7h)}{i2}
\fmfforce{(.5w,.3h)}{i4}
\end{fmfgraph*}
}}
&
\large{b)}
\scriptsize{
\fmfframe(0,0)(0,-30){
\begin{fmfgraph*}(120,120)
\fmfleft{v1}
\fmfright{v2}
\fmf{plain}{v1,i1}
\fmf{plain}{i4,v2}
\fmf{plain,left=0.2,label=1}{i1,i2}
\fmf{plain,left=0.2,label=4}{i3,i4}
\fmf{plain,right,label=5}{i1,i4}
\fmffreeze
\fmf{plain,left=0.5,label=2}{i2,i3}
\fmf{plain,right=0.5,label=3}{i2,i3}
\fmfforce{(.35w,.65h)}{i2}
\fmfforce{(.65w,.65h)}{i3}
\end{fmfgraph*}
}}
\\
\large{c)}
\scriptsize{
\fmfframe(0,0)(0,-30){
\begin{fmfgraph*}(120,120)
\fmfleft{v1}
\fmfright{v2}
\fmf{plain}{v1,i1}
\fmf{plain}{i3,v2}
\fmf{plain,left=0.5,label=1}{i1,i2}
\fmf{plain,left=0.5,label=3}{i2,i3}
\fmf{plain,right,label=4}{i1,i3}
\fmffreeze
\fmfforce{(.5w,.7h)}{i2}
\fmfforce{(.3w,.5h)}{i1}
\fmfforce{(.7w,.5h)}{i3}
\fmfforce{(.5w,.9h)}{i4}
\fmf{plain,right}{i2,i4,i2}
\fmfv{label=2}{i4}
\end{fmfgraph*}
}}
&
\large{d)}
\scriptsize{
\fmfframe(0,0)(0,-30){
\begin{fmfgraph*}(120,120)
\fmfleft{v1}
\fmfright{v2}
\fmf{plain}{v1,i1}
\fmf{plain}{i1,v2}
\fmf{plain,left=0.7,label=1}{i1,i2}
\fmf{plain,left,label=2}{i2,i3}
\fmf{plain,right,label=3}{i2,i3}
\fmf{plain,left=0.7,label=4}{i3,i1}
\fmffreeze
\fmfforce{(.4w,.8h)}{i2}
\fmfforce{(.6w,.8h)}{i3}
\end{fmfgraph*}
}}
\\
\large{e)}
\scriptsize{
\fmfframe(0,0)(0,-30){
\begin{fmfgraph*}(120,120)
\fmfleft{v1}
\fmfright{v2}
\fmf{plain}{v1,i1}
\fmf{plain}{i1,v2}
\fmf{plain,left,label=1}{i1,i2}
\fmf{plain,left}{i2,i3,i2}
\fmfv{label=2,label.dist=0.02w}{i3}
\fmf{plain,left,label=3}{i2,i1}
\fmffreeze
\fmfforce{(.5w,.75h)}{i2}
\fmfforce{(.5w,.9h)}{i3}
\end{fmfgraph*}
}}&
\large{f)}
\scriptsize{
\fmfframe(0,0)(0,-30){
\begin{fmfgraph*}(120,120)
\fmfleft{v1}
\fmfright{v2}
\fmf{plain}{v1,i1}
\fmf{plain}{i2,v2}
\fmf{plain,left,label=1}{i1,i2}
\fmf{plain,right=0.5,label=2}{i1,i3}
\fmf{plain,left=0.5,label=3}{i1,i3}
\fmf{plain,right=0.5,label=4}{i3,i2}
\fmffreeze
\fmfforce{(.5w,.25h)}{i3}
\fmfforce{(.25w,.5h)}{i1}
\fmfforce{(.75w,.5h)}{i2}
\end{fmfgraph*}
}}
\\
\large{g)}
\scriptsize{
\fmfframe(0,0)(0,-30){
\begin{fmfgraph*}(120,120)
\fmfleft{v1}
\fmfright{v2}
\fmf{plain}{v1,i1}
\fmf{plain}{i2,v2}
\fmf{plain,left,label=1}{i1,i2}
\fmf{plain,label=2}{i1,i2}
\fmf{plain,right,label=3}{i1,i2}
\fmfforce{(.25w,.5h)}{i1}
\fmfforce{(.75w,.5h)}{i2}
\end{fmfgraph*}
}}
&
\large{h)}
\scriptsize{
\fmfframe(0,0)(0,-30){
\begin{fmfgraph*}(120,120)
\fmfleft{v1}
\fmfright{v2}
\fmf{plain}{v1,i1}
\fmf{plain}{i3,v2}
\fmf{plain,left,label=1}{i1,i2}
\fmf{plain,right,label=2}{i1,i2}
\fmf{plain,left,label=3}{i2,i3}
\fmf{plain,right,label=4}{i2,i3}
\fmfforce{(.2w,.5h)}{i1}
\fmfforce{(.8w,.5h)}{i3}
\end{fmfgraph*}
}}
\end{tabular}}
\caption{\label{topose}
Topologies of two loop two point diagrams}
\end{figure}
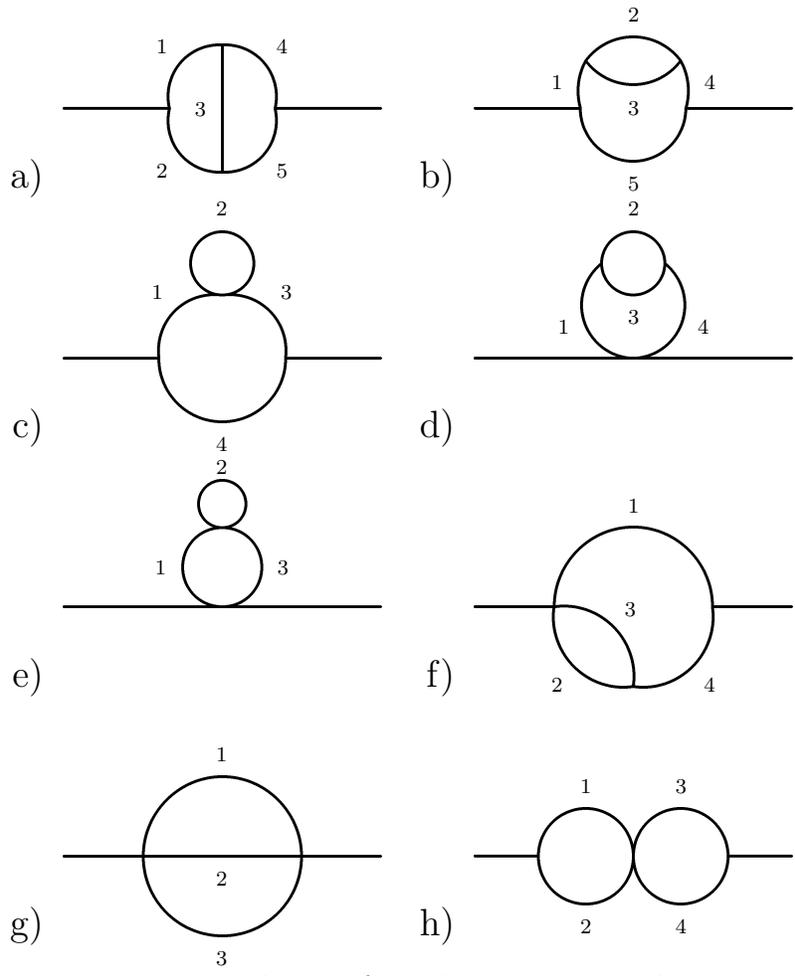

Depending on the number of fields and the specific form of the Lagrangian
the number of possible diagrams may vary. However, there is only a
restricted set of different topologies these diagrams may have.
At one loop level there are only two possible topologies. These
two are the diagrams $\alpha$ and $\beta$ shown in Fig. \ref{topo1l}. 
At two loop level the number of different
topologies is eight. These are given in Fig. \ref{topose}.
\begin{fmffile}{topot}

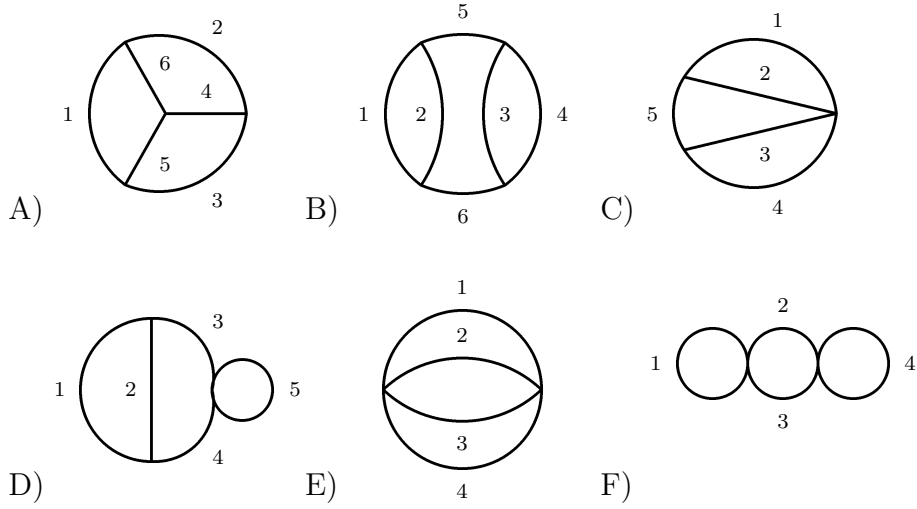
\begin{figure}
\centering{
\begin{tabular}{lll}
\normalsize{A)}
\scriptsize{
\fmfframe(10,10)(10,10){
\begin{fmfgraph*}(60,60)
\fmfsurround{v1,v2,v3}
\fmf{plain,label=4}{v1,i1}
\fmf{plain,label=6}{v2,i1}
\fmf{plain,label=5}{v3,i1}
\fmf{plain,right=0.5,label=2}{v1,v2}
\fmf{plain,right=0.5,label=1}{v2,v3}
\fmf{plain,right=0.5,label=3}{v3,v1}
\end{fmfgraph*}}}
&\normalsize{B)}
\scriptsize{
\fmfframe(10,10)(10,10){
\begin{fmfgraph*}(60,60)
\fmfsurround{v0,v1,v2,v00,v3,v4}
\fmf{plain,left=0.5,label=4}{v1,v4}
\fmf{plain,left=0.3,label=2,l.s=right}{v2,v3}
\fmf{plain,right=0.3,label=3,l.s=left}{v1,v4}
\fmf{plain,right=0.5,label=1}{v2,v3}
\fmf{plain,right=0.2,label=5}{v1,v2}
\fmf{plain,right=0.2,label=6}{v3,v4}
\end{fmfgraph*}}}
&\normalsize{C)}
\scriptsize{
\fmfframe(10,10)(10,10){
\begin{fmfgraph*}(60,60)
\fmfsurround{v3,v0,v00,v1,v2,vo,voo}
\fmf{plain,left=0.7,label=1}{v1,v3}
\fmf{plain,label=3}{v2,v3}
\fmf{plain,label=2,label.side=left}{v1,v3}
\fmf{plain,right=0.7,label=4}{v2,v3}
\fmf{plain,right=0.3,label=5}{v1,v2}
\end{fmfgraph*}}}
\\
\normalsize{D)}
\scriptsize{
\fmfframe(20,10)(0,10){
\begin{fmfgraph*}(60,60)
\fmfsurround{v1,v2,v3}
\fmfv{label=5}{v1}
\fmf{plain,right}{v1,i1}
\fmf{plain,left}{v1,i1}
\fmf{plain,right=0.5,label=3}{i1,v2}
\fmf{plain,right=0.5,label=4}{v3,i1}
\fmf{plain,right,label=1}{v2,v3}
\fmf{plain,label=2}{v2,v3}
\end{fmfgraph*}}}
&\normalsize{E)}
\scriptsize{
\fmfframe(10,10)(10,10){
\begin{fmfgraph*}(60,60)
\fmfsurround{v1,v2}
\fmf{plain,right,label=1}{v1,v2}
\fmf{plain,left,label=4}{v1,v2}
\fmf{plain,right=0.4,label=2}{v1,v2}
\fmf{plain,left=0.4,label=3}{v1,v2}
\end{fmfgraph*}}}
&\normalsize{F)}
\scriptsize{
\fmfframe(10,10)(10,10){
\begin{fmfgraph*}(80,80)
\fmfsurround{v1,v2}
\fmf{plain,right}{v1,i1}
\fmf{plain,left}{v1,i1}
\fmf{plain,right,label=2}{i1,i2}
\fmf{plain,left,label=3}{i1,i2}
\fmf{plain,right}{i2,v2}
\fmf{plain,left}{i2,v2}
\fmfv{label=1}{v2}
\fmfv{label=4}{v1}
\end{fmfgraph*}}}
\end{tabular}
}

\caption{\label{topoep}Topologies of three loop vacuum diagrams}
\end{figure}

\end{fmffile}

For the effective potential one needs the vacuum diagrams 
\cite{ref:Jackiw}. At one loop
level there is of course only one possible topology, and two at the
two loop level. At the three loop level the number is six. These diagrams are
shown in Fig. \ref{topoep}.

Some of the integrals corresponding to these diagrams do not converge, but
when they do, there is a relation between vacuum and two point integrals.

\begin{lem}
\label{rellemma}
Let $f_2(p)$ be a two point integral with an external momentum of $p$ and
$F_0(m)$ the vacuum diagram obtained by connecting the outer legs of $f_2$
with a particle of mass $m$. Then the following relation holds:
\beq
f_2(p)={2\pi \I\over p}\left (F_0(\I p)-F_0(-\I p)\right).
\label{lemmaequ}
\eeq
\end{lem}

The lemma can easily be proven by taking a Fourier transform of both sides
of Eq. (\ref{lemmaequ}).

One should also notice that in case there are more than one propagator
with the same momentum, they can be separated to two integrals by
partial fractioning or be written as a derivative:
\bea
\label{partfrac}
{1\over (p^2+m_1^2)(p^2+m_2^2)}&=&{1\over m_2^2-m_1^2}\left({1\over p^2+m_1^2}
-{1\over p^2+m_2^2}\right),\\
\label{deriv}
{1\over (p^2+m^2)^2}&=&-{1\over 2m}{\partial\over\partial m}{1\over p^2+m^2}.
\eea

Some of the integrals also factorize into separate parts. Using these relations
the number of integrals to be calculated can be reduced to six: $\alpha$, 
$\beta$, g, A, C and E. Of these, diagrams $\alpha$ and $\beta$ are easily 
evaluated. Diagram 
$\gamma$ is only a special case of g. 
Diagrams 
a, b and f
are related to diagrams A, B and C, respectively, by Lemma \ref{rellemma}.
Finally, all the diagrams c, d, e, h, D and F 
factorize to products of simpler diagrams. The results for all the integrals
are given in explicit form in appendix \ref{appint}.

\subsection{Diagram g}

Consider now the diagram g called the sunset. Diagram
$\gamma$ 
is a special case of this with vanishing external momentum. Let the masses
of the particles be $m_1$, $m_2$ and $m_3$. In this integral
the coordinate space method \cite{ref:arnoldesp,ref:fourier} will be used.
In $\dim$ dimensions the Fourier transform of the propagator is
\beq
V_i(\vec x)=(\pi\mu^2)^\ve{1\over(2\pi)^{{3\over 2}-\ve}}
\left({m_i\over x}\right)^{{1\over 2}-\ve}K_{{1\over 2}-\ve}(m_ix).
\eeq
Here $K_\nu(x)$ is the modified Bessel function. After the transform the
integral reads
\beq
\In_g(k;m_1,m_2,m_3)=\mu^{2\ve}\int d^\dim R 
\e^{i\vec k\cdot\vec R}\prod_iV_i(\vec R).
\eeq
The divergence occurs only on the limit $R\rightarrow 0$. Thus the integration
can be split
at $R=r$:
\bea
\In_g(k)&=&\left({\e^\gamma\bar\mu^2\over2k}\right)^{-\ve}\int_0^rdR
R^{{3\over 2}-\ve}
J_{{1\over 2}-\ve}(kR)\prod_iV_i(R)\nn\\
&&
+{4\pi\over k}\int_r^\infty dRR\sin(kR)\prod_iV_i(R)\nn\\
&\equiv&\In_g^{(a)}(k)+\In_g^{(b)}(k),
\eea
where $J_\nu(x)$ is the Bessel function and $\bar\mu$ is the 
$\overline{\rm MS}$ scale parameter $\bar\mu^2=\e^{-\gamma}4\pi\mu^2$.

Since $\In_g^{(b)}$ converges, the 
usual three dimensional Fourier transform of the propagator can be used.
In $\In_g^{(a)}$, 
that is when $R<r$, one can approximate the Bessel functions by 
the lowest order terms in their Laurent series: 
\bea
J_{{1\over 2}-\ve}(kR)
&=&{1\over \Gamma({3\over 2}-\ve)}\biggl({1\over 2}kR\biggr)
^{{1\over 2}-\ve}+{\cal O}\left((kR)^{3\over 2}\right),\\
V_i(R)&=&\left({\e^\gamma\bar\mu^2\over 4}\right)^\ve
{\Gamma({1\over 2}-\ve)\over\Gamma({1\over 2})}{1\over 4\pi}R^{-1+2\ve}\nn\\
&&-(\e^\gamma\bar\mu^2)^\ve{\Gamma(-{1\over 2}+\ve)\over \Gamma(-{1\over 2})}
{1\over 4\pi}m_i^{1-2\ve}
+{\cal O}(R).
\eea
The error vanishes as the limit 
$r\rightarrow 0$ is taken.
For the present integral only the ${\cal O}(R^{-1})$ term of $V_i$ is needed.
Now one is left with only a straightforward task of integrating over
powers of $R$. All the terms that are singular at the lower limit
$R=0$ are treated by analytical continuation to sufficiently great
values of $\ve$ so that they vanish. 

The result is
\beq
\In_g^{(a)}(k)={1\over(4\pi)^2}\left({1\over 4\ve}+\log\bar\mu r
+{1\over 2}+\gamma\right).
\eeq

$\In_g^{(b)}$ is a normal three dimensional integral and it can be calculated 
using normal methods of multidimensional integration:
\bea
\In_g^{(b)}(k)&=&{1\over(4\pi)^2}\left(1-{m_1+m_2+m_3\over k}
\arctan{k\over m_1+m_2+m_3}\right.\nn\\
&&
\left.-\gamma-\log r-{1\over 2}\log\left((m_1+m_2+m_3)^2+k^2\right)\right).
\eea

Now one can write down the result:
\bea
\In_g(k)&=&{1\over(4\pi)^2}\left({1\over 4\ve}-{m_1+m_2+m_3\over k}
\arctan{k\over m_1+m_2+m_3}\right.\nn\\
&&\left.
+{1\over 2}\log{\bar\mu^2\over(m_1+m_2+m_3)^2+k^2}+{3\over 2}\right).
\eea

\subsection{Reducible diagrams}
The integral g together with the well known results for the integrals 
$\alpha$ and $\beta$ make it possible to calculate all the
reducible integrals
using Lemma \ref{rellemma} and Eq.
(\ref{partfrac}).
As a simple example, consider the diagram c:  
\bea
\In_c&=&\int\meas{k}\meas{q}\nn\\
&&
{1\over (k^2+m_1^2)(q^2+m_2^2)(k^2+m_3^2)((p-k)^2+m_4^2)}.
\eea
The $q$-integration is nothing but $\In_\alpha$. One then uses Eq. 
(\ref{partfrac}) to obtain
\bea
\In_c&=&-{m_2\over 4\pi}{1\over m_3^2-m_1^2}
\int\meas{k}
\biggl({1\over k^2+m_1^2}-{1\over k^2+m_3^2}\biggr)
{1\over (p-k)^2+m_4^2}\nn\\
&=&{m_2\over 4\pi}{1\over m_1^2-m_3^2}\left(
\In_\beta(m_1,m_4)-\In_\beta(m_3,m_4)\right)\nn\\
&=&
\label{Ic}
{m_2\over (4\pi)^2p(m_1^2-m_3^2)}\left(
\arctan{p\over m_1+m_4}
-\arctan{p\over m_3+m_4}\right).
\eea

In a similar way the diagram D can be factorized:
\bea
\In_D&=&\In_\alpha(m_5){1\over m_4^2\!-\!m_3^2}
\left(\In_\gamma(m_1,m_2,m_3)-\In_\gamma(m_1,m_2,m_4)\right)\nn\\
&=&
{m_5\over (4\pi)^3(m_3^2-m_4^2)}\log{m_1+m_2+m_4\over m_1+m_2+m_3}.
\eea

The results of the integrals are convergent, but the integrals themselves
are actually not, since they both contain the simple loop $\alpha$, which
is divergent. In dimensional regularization this divergence vanishes,
but Lemma \ref{rellemma} not proven to hold with dimensional regularization.
However, this loop is in both cases factorized as a separate
integral  and the
remaining other factor is convergent. Therefore Lemma \ref{rellemma}
should hold for these diagrams:
\bea
\lefteqn{\In_c(p;m_1,m_2,m_3,m_4)=}\nn\\
&&
{2\pi \I\over p}\left(\In_D(\I p,m_4,m_1,m_3,m_2)
-\In_D(-\I p,m_4,m_1,m_3,m_2)\right).
\eea

Substitute the previous result to the right hand side to obtain
\bea
\In_c
&=&
{m_2\over (4\pi)^2 p(m_1^2-m_3^2)}
\left(\arctan{p\over m_1+m_4}
-\arctan{p\over m_3+m_4}\right),
\eea
which is the same as Eq. (\ref{Ic}).

\subsection{Diagram E}

Since the integrals E and g 
do not converge, Lemma \ref{rellemma} does not hold. Therefore one must 
start the calculation of E from the beginning. The 
coordinate space integral is
\beq
\In_E(m_1,m_2,m_3,m_4)=\mu^{2\ve}\int d^\dim R\prod_{n=1}^4 V_i(\vec R).
\eeq

Just like before, this is separated to two parts, only one of which is 
divergent:
\bea
\In_E&=&\biggl({\e^\gamma\bar\mu^2\over 4}\biggr)^{-\ve}
{\Gamma({3\over 2})\over \Gamma({3\over 2}-\ve)}
4\pi\int_0^rdRR^{2-2\ve}\prod_iV_i(R)\nn\\
&&
+{1\over (4\pi)^3}\int_r^\infty dRR^2 {1\over R^4}\e^{-\sum_im_iR}.
\eea

In the expansion of $V_i$ one now has to take into account also the term 
of order ${\cal O}(1)$. Then the first integral reads
\bea
\In_E^{(a)}&=&{1\over (4\pi)^3}\left({\e^\gamma\bar\mu^2\over 4}\right)^{-\ve}
{\Gamma({3\over 2})\over \Gamma({3\over 2}-\ve)}
\int_0^rdRR^{2-2\ve}\nn\\
&&
\left[
\left({\e^\gamma\bar\mu^2\over 4}\right)^{4\ve}
\left({\Gamma({1\over 2}-\ve)\over\Gamma({1\over 2})}\right)^4
R^{-4+8\ve}\right.\nn\\
&&
\left.-
\left({\e^\gamma\bar\mu^2\over 4}\right)^{3\ve}
\left({\Gamma({1\over 2}-\ve)\over\Gamma({1\over 2})}\right)^3
(\e^\gamma\bar\mu^2)^\ve
{\Gamma(-{1\over 2}+\ve)\over\Gamma(-{1\over 2})}
R^{-3+6\ve}\sum_im_i^{1-2\ve}\right].
\eea

This simple integral gives
\bea
\In_E^{(a)}&=&
{1\over (4\pi)^3}\biggl[
-{1\over r}-{1\over 4}\sum_im_i\left(
{1\over\ve}+4+4\gamma+2\log{r^2\bar\mu^3\over2m_i}\right)\biggr].
\eea

$\In_E^{(b)}$ is also easily evaluated and one can write the result for the
whole integral:
\beq
\In_E=
{1\over (4\pi)^3}\sum_{i=1}^4m_i
\left(-{1\over 4\ve}+2+{1\over 2}\log{2m_i\over \bar\mu}
+\log{\sum_jm_j\over\bar\mu}\right)\, .
\eeq

\subsection{Diagram C}

Consider now the diagram C. This integral converges and Fourier transform
can be used directly in three dimensional space. The masses are as shown
in Fig.
\ref{topoep}.
\bea
\In_C&=&\int_{p,q,k}{1\over (p^2+m_1^2)((k-p)^2+m_2^2)((k-q)^2+m_3^2)}\nn\\
&&{1\over(q^2+m_4^2)
(k^2+m_5^2)}\, .
\eea

Taking Fourier transform of this gives
\beq
\label{fourJ}
\In_C={1\over (4\pi)^5}\int d^3x_1{\e^{-(m_1+m_2)x_1}\over x_1^2}
\int d^3x_2{\e^{-m_5|x_1+x_2|-(m_3+m_4)x_2}\over |x_1+x_2|x_2^2}.
\eeq

Concentrate now on the latter integration
\bea
\lefteqn{\int d^3x_2{\e^{-m_5|x_1+x_2|-(m_3+m_4)x_2}\over |x_1+x_2|x_2^2}}\nn\\
&=&{2\pi\over m_5x_1}\left[\e^{-m_5x_1}\int_0^{x_1}{dx\over x}
(\e^{-(m_3+m_4-m_5)x}-\e^{-(m_3+m_4+m_5)x})\right.\nn\\
&&
\left.+(\e^{m_5x_1}-\e^{-m_5x_1})\int_{x_1}^\infty{dx\over x}
\e^{-(m_3+m_4+m_5)x}\right]
\nn\\
&=&{2\pi\over m_5x_1}\biggl[\e^{-m_5x_1}\left(\log{m_3+m_4+m_5
\over m_3+m_4-m_5}
+{\rm Ei}\left((m_3+m_4-m_5)x_1\right)\right)\nn\\
&&
-\e^{m_5x_1}{\rm Ei}((m_3+m_4+m_5)x_1)\biggr],
\eea
where $\Ei(x)$ is the exponential integral.

Substituting this into Eq. (\ref{fourJ}) and integrating gives the result
\bea
\In_C&=&\lim_{r\rightarrow 0}{1\over (4\pi)^3 2 m_5}
\biggl[\log{m_3+m_4+m_5\over m_3+m_4-m_5}
\left(-\gamma-\log(m_1+m_2+m_5)r\right)\nn\\
&&
-{1\over 2}\left(\zeta(2)+\left(\gamma+\log(m_3+m_4-m_5)r\right)^2\right)
-{\rm Li_2}\left(-{m_1+m_2+m_5\over m_3+m_4-m_5}\right)\nn\\
&&
+{1\over 2}\left(\zeta(2)+\left(\gamma+\log(m_3+m_4+m_5)r\right)^2\right)
+{\rm Li_2}\left(-{m_1+m_2-m_5\over m_3+m_4+m_5}\right)\biggr]\nn\\
&=&{1\over (4\pi)^3 2 m_5}
\biggl[\log{m_3+m_4+m_5\over m_3+m_4-m_5}
\log{\sqrt{(m_3+m_4)^2-m_5^2}\over m_1+m_2+m_5}\nn\\
&&
+{\rm Li_2}\left(-{m_1+m_2-m_5\over m_3+m_4+m_5}\right)
-{\rm Li_2}\left(-{m_1+m_2+m_5\over m_3+m_4-m_5}\right)
\biggr].
\eea
${\rm Li_2}(x)$ is the Euler dilogarithm function and $\zeta(x)$ is the
Riemann zeta function. 

\section{Mercedes and the master integral}
\label{mastersect}
\begin{fmffile}{topomb}
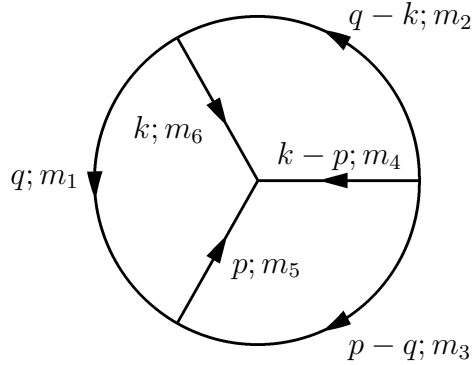
\begin{figure}
\centering{
\fmfframe(10,10)(10,10){
\begin{fmfgraph*}(120,120)
\fmfsurround{v1,v2,v3}
\fmf{plain_arrow,label=$k-p; m_4$}{v1,i1}
\fmf{plain_arrow,label=$k; m_6$}{v2,i1}
\fmf{plain_arrow,label=$p; m_5$}{v3,i1}
\fmf{plain_arrow,right=0.58,label=$q-k; m_2$}{v1,v2}
\fmf{plain_arrow,right=0.58,label=$q; m_1$}{v2,v3}
\fmf{plain_arrow,left=0.58,label=$p-q; m_3$}{v1,v3}
\end{fmfgraph*}}
}

\caption{\label{mersuimpulssit}
The momenta and the masses of the Mercedes diagram}
\end{figure}

\end{fmffile}

\subsection{Mercedes integral}
The last diagram to be considered is A. 
Many different techniques have been proposed for the calculation
%\cite{ref:series,ref:series2,ref:Mellin,ref:Gegenbauer,ref:Scharf,%
%ref:Bauberger,ref:Kreimer,ref:Ghinculov,ref:Avdeev,ref:Broadhurst},
[10--19],
most often in four dimensions.
The 
coordinate space method of the preceeding subsection does not work
well even though the integral converges and three dimensional transform can
be used. The transformed integral is namely as complicated as the
original one. Other approachs are based on some series expansion of the 
integral
but then the resummation
of the series is usually not possible.
Many other methods, like Mellin-Barnes transformation
 and Gegenbauer
polynomials, work well only in four dimensional space.
In four dimensions one has been able to reduce the general combination of
masses to a one-dimensional integral.

The approach to be used here is similar to that of Kotikov
\cite{ref:Kotikov}. A differential equation which
the integral must satisfy is constructed and then this equation is solved.
For propagators a
shorthand notation will be used:
\beq
\Delta^i_p={1\over p^2+m_i^2}.
\eeq

With the masses and the momenta defined as in Fig. \ref{mersuimpulssit},
the integral reads
\bea
\In_A(m_1,m_2,m_3,m_4,m_5,m_6)
&=&\int \Delta_q^1\Delta_{q-k}^2\Delta_{p-q}^3\Delta_{k-p}^4\Delta_p^5
\Delta_k^6\nn\\
&
\equiv&
\int \Delta^{123456}.
\eea
Here integration over all momenta is assumed.

The first step is to integrate this by parts. The boundary terms vanish.
\bea
\In_A&=&-{1\over 3}\int q_i{\partial\over\partial q_i}
\Delta^{123456}\nn\\
&=&-{1\over 3}\int \left( 
-2q^2\Delta^1_q
-2\vec{q}\cdot(\vec{q}-\vec{k})\Delta^2_{q-k}
-2\vec{q}\cdot(\vec{q}-\vec{p})\Delta^3_{p-q}
\right)\Delta^{123456}.
\eea

Using conservation of momentum and then simplifying one gets the result
\bea
\In_A&=&\int \left(
2m_1^2\Delta^1_q-(\Delta^1_q)^{-1}\Delta_{q-k}^2
+(\Delta^6_k)^{-1}\Delta_{q-k}^2
+(m_1^2+m_2^2-m_6^2)\Delta_{q-k}^2\right.\nn\\
&&
\left.-(\Delta^1_q)^{-1}\Delta^3_{p-q}+(\Delta^5_p)^{-1}\Delta^3_{p-q}
+(m_1^2+m_3^2-m_5^2)\Delta^3_{p-q}
\right)\Delta^{123456}\nn\\
&=&2m_1^2 K_2-J_2^1+J_2^6+(m_1^2+m_2^2-m_6^2)K_2\nn\\
&&-J_3^2+J_3^5+(m_1^2+m_3^2-m_5^2)K_3.
\label{linear}
\eea
Here $K_i$ is $\In_A$ with the propagator of particle $i$ squared and 
$J_i^j$ is a diagram of type $\In_C$ which has been constructed by removing
the propagator of $j$ from $\In_A$ with propagator $i$ squared.
Then by Eq. (\ref{deriv}) one has 
\beq
K_i=-{1\over 2m_i}{\partial\over\partial m_i}\In_A(m_1,m_2,m_3,m_4,m_5,m_6).
\eeq
$J_i^j$ can be obtained from the following expression by substituting the
corresponding masses
$$
\tilde\In_C(m_1,m_2,m_3,m_4,m_5)\equiv
-{1\over 2m_1}{\partial\over\partial m_1}\In_C(m_1,m_2,m_3,m_4,m_5)
$$
\beq
={1\over(4\pi)^32m_1
((m_1+m_2)^2-m_5^2)}
\log{m_1+m_2+m_3+m_4\over m_3+m_4+m_5}.
\label{jtilde}
\eeq
In appendix \ref{appint} the explicit expressions are
given for necessary $J_i^j$.
Equation (\ref{linear}) can be transformed to a more suitable form
by using the symmetry properties of the diagram. Obviously, the diagram
is invariant in the following permutations of masses:
\beq
\pmatrix{1&2&3\cr 4&5&6} \Rightarrow \pmatrix{2&3&1\cr 5&6&4} \Rightarrow
\pmatrix{3&1&2\cr 6&4&5},
\eeq
because they are equivalent to rotations of the diagram.
 
Using these symmetries, Eq. (\ref{linear}) 
can be written in matrix form
\bea
\M(m_1,m_2,m_3,m_4,m_5,m_6)K\equiv&&\nn\\
\pmatrix{2m_1^2&A&B\cr A&2m_2^2&C\cr B&C&2m_3^2}\pmatrix{K_1\cr K_2\cr K_3}
&=&\pmatrix{\In_A+J_1\cr \In_A+J_2\cr \In_A+J_3}\equiv \In_A I+J,
\eea
where $I=(1,1,1)^T$ and $J=(J_1,J_2,J_3)^T$,
\bea
A&=&m_1^2+m_2^2-m_6^2\nn\\
B&=&m_1^2+m_3^2-m_5^2\nn\\
C&=&m_2^2+m_3^2-m_4^2
\eea
and
\bea
\label{Jdef}
J_1&=&J_2^1-J_2^6+J_3^1-J_3^5\nn\\
J_2&=&J_3^2-J_3^4+J_1^2-J_1^6\nn\\
J_3&=&J_1^3-J_1^5+J_2^3-J_2^4.
\eea
However, since in the following all masses except $m_1$ will be fixed and
it occurs in the matrix only in quadratic form, the following notation will
be used for simplicity:
\beq
\M(m_1^2)\equiv\M(m_1,m_2,m_3,m_4,m_5,m_6).
\label{Mnotaatio}
\eeq

\subsection{Degenerate case}

Consider first the degenerate case. If $|\M(m_1^2)|=0$ the matrix has an 
eigenvalue of zero. Let $U$ be the
corresponding eigenvector. Then
\beq
0=U^T\M(m_1^2)K=U^TI\In_A+U^TJ.
\eeq

Let $D_i$ be the minor of the determinant of matrix $\M(m_1^2)$
with one of the rows and column $i$ removed.
Then a suitable eigenvector is
\beq
U=\pmatrix{D_1\cr -D_2\cr D_3}.
\eeq

Hence, the result for $\In_A$ can be written down:
\beq
\In_A=-{U^TJ\over U^TI}=-{|\M_J|\over|\M_I|},
\eeq
where $\M_X$ denotes matrix $\M(m_1^2)$ with one of the rows or columns 
replaced by vector
$X$. Not every choice of row or column is always possible, but one can
use the symmetries of the diagram to transform the matrices to a form in
which the chosen row is the first one. Therefore the first row will be used
here.

As an example let us calculate the special case $m_1=0$,\ $m_5=m_3$,\
$m_6=m_2$.
Then the matrix $\M$ reads
\beq
\M=\pmatrix{0&0&0\cr
            0&2m_2^2&m_2^2+m_3^2-m_4^2\cr
            0&m_2^2+m_3^2-m_4^2&2m_3^2\cr}.
\eeq
Then
\bea
\In_A&=&-{\left|\matrix{J_1&J_2&J_3\cr
                 0&2m_2^2&m_2^2+m_3^2-m_4^2\cr
                 0&m_2^2+m_3^2-m_4^2&2m_3^2\cr}\right|
\over
    \left|\matrix{1&1&1\cr
                 0&2m_2^2&m_2^2+m_3^2-m_4^2\cr
                 0&m_2^2+m_3^2-m_4^2&2m_3^2\cr}\right|}=-J_1\nn\\
&=&-J_2^1+J_2^6-J_3^1+J_3^5\nn\\
&=&-\tilde\In_C(m_2,m_3,m_2,m_3,m_4)+\tilde\In_C(m_2,m_4,0,m_3,m_3)\nn\\
&&-\tilde\In_C(m_3,m_2,m_3,m_2,m_4)+\tilde\In_C(m_3,m_4,0,m_2,m_2).
\eea

Substituting the expression (\ref{jtilde}) one obtains
\bea
\lefteqn{\In_A(0,m_2,m_3,m_4,m_3,m_2)=}\nn\\
&&{1\over (4\pi)^3}{1\over 2}\Biggl[
{\log{2m_2\over m_2+m_3+m_4}\over m_3(m_2^2-(m_3^2+m_4^2))}+
{\log{2m_3\over m_2+m_3+m_4}\over m_2(m_3^2-(m_2^2+m_4^2))}\nn\\
&&
+\left({1\over m_2}+{1\over m_3}\right)
{\log{2(m_2+m_3)\over m_2+m_3+m_4}\over m_4^2-(m_2^2+m_3^2)}\Biggr].
\eea

In the case $m_2=m_3=m$, $m_4=0$ this result simplifies to
\beq
\In_A(0,m,m,0,m,m)={1\over (4\pi)^3}{1\over 4 m^3}(1-\log 2),
\eeq
which is the result of \cite{ref:Zhai}.

\subsection{Non-degenerate case}

Consider now the non-degenerate case $|\M|\neq 0$. Now the matrix is
invertible. The equation can then be solved for $K_1$. This leads to
a first order linear differential equation:
\beq
\label{diffeq}
{\partial\over\partial m_1^2}\In_A=-K_1=
-{|\M_I|\In_A+|\M_J|\over |\M(m_1^2)|},
\eeq
where in $\M_X$ substitution of $X$ to the first row is assumed. 
From now on for
$\In_A$ a notation similar to (\ref{Mnotaatio}) will be used:
\beq
\In_A(m_1^2)\equiv\In_A(m_1,m_2,m_3,m_4,m_5,m_6).
\eeq

In the domain where the coefficients on the right hand side are continuous
(\ref{diffeq}) now has a unique solution. If $I_A$ is this solution,
it has a form
\bea
\label{uniquesol}
I_A(m_1^2)&=&-\exp\biggl(-\int^{m_1^2}{|\M_I(t)|\over|\M(t)|}dt\biggr)\nn\\
&&
\Biggl[ \int^{m_1^2}
\exp\biggl(-\int^s{|\M_I(t)|\over|\M(t)|}dt\biggr)
{|\M_J(s)|\over|\M(s)|}ds+C\Biggr].
\eea

The value of the integration constant $C$ is to be determined. Let the
lower limit of the integrations be $\tilde m^2$,
a point such that the coefficients are
continuous. Now let $m_1^2$ approach $\tilde m^2$ to see that 
in this case $C=\In_A(\tilde m^2)$, since the exponential factor
approaches unity and the integral in the brackets vanishes.
Now let $m_0$ be such that $|\M(m_0^2)|=0$. 
Since $I_A(m_1^2)$ coincides with $\In_A(m_1^2)$
when $m_1^2<m_0^2$ and $\In_A$ is continuous also when $m_1^2=m_0^2$,
they must coincide also at the point $m_1^2=m_0^2$ for $I_A$ to be
continuous.
Thus, when $m_1^2\leq m_0^2$,
\beq
\In_A(m_1^2)=I_A(m_1^2).
\eeq
However, when $m_1^2\rightarrow m_0^2$, the exponential factor in 
(\ref{uniquesol}) diverges.
Therefore the expression inside the
brackets must vanish and one obtains
\beq
\In_A(m_1^2)=-\int_{m_0^2}^{m_1^2}\exp
\biggl(\int_{m_1^2}^s{|\M_I(t)|\over|\M(t)|}dt\biggr)
{|\M_J(s)|\over|\M(s)|}ds.
\eeq

Now the following relation is true:
\beq
|\M_I(m_1^2)|={1\over 2}{\partial\over\partial m_1^2}|\M(m_1^2)|.
\eeq
This can be seen as follows. Consider the determinant expanded as a sum of
minors:
\beq
|\M(m_1^2)|=2m_1^2\left|\matrix{2m_2^2&C\cr C&2m_3^2}\right|
-A\left|\matrix{A&C\cr B&2m_3^2}\right|
+B\left|\matrix{A&2m_2^2\cr B&C}\right|.
\eeq
The first minor is constant in $m_1^2$ and 
$\partial A/\partial m_1^2=\partial B/\partial m_1^2=1$. Therefore 
\bea
{\partial\over\partial m_1^2}|\M(m_1^2)|
&=&2\left|\matrix{2m_2^2&C\cr C&2m_3^2}\right|
-\left|\matrix{A&C\cr B&2m_3^2}\right|
+B\left|\matrix{A&2m_2^2\cr B&C}\right|\nn\\
&&-A\left|\matrix{1&C\cr 1&2m_3^2}\right|
+B\left|\matrix{1&2m_2^2\cr 1&C}\right|\nn\\
&=&\left|\matrix{1&1&1\cr A&2m_2^2&C\cr B&C&2m_3^2}\right|
+\left|\matrix{1&A&B\cr 1&2m_2^2&C\cr 1&C&2m_3^2}\right|\nn\\
&=&2\left|\matrix{1&1&1\cr A&2m_2^2&C\cr B&C&2m_3^2}\right|=2|\M_I(m_1^2)|.
\eea

Thus, the integral in the exponential can be calculated and one obtains
\bea
\In_A(m_1^2)&=&-\int_{m_0^2}^{m_1^2}\sqrt{|\M(s)|\over|\M(m_1^2)|}
{|\M_J(s)|\over|\M(s)|}ds\nn\\
&=&-{1\over\sqrt{|\M(m_1^2)|}}\int_{m_0^2}^{m_1^2}{|\M_J(s)|\over
\sqrt{|\M(s)|}}ds
\nn\\
&=&-{2\over\sqrt{|\M(m_1^2)|}}\int_{m_0}^{m_1}{|\M_J(x^2)|\over
\sqrt{|\M(x^2)|}}xdx.
\label{mersutulos}
\eea

This result holds only if the integrand has no singularities between
$m$ and $m_0$. 
However, $\M_J$ diverges only if $m_1=0$, $m_2=0$ or $m_3=0$, i.e. there
is no closed massive loop in the diagram. That case must be treated
separately and leads to infrared divergences.
An appropriate choice of $m_0$ ensures that 
the denominator $\sqrt{\M(x^2)}$ has no zeros on the domain of
integration.

When evaluating the integral (\ref{mersutulos}) one needs the following
integral
\beq
\int_0^1{\log(1+\alpha t)dt\over\sqrt{1-\beta t^2+\gamma t^4}(1+\delta t)}.
\eeq
Unfortunately this integral is not expressible in terms of usual 
special functions. Therefore one
needs numerical evaluation of the last expression. Since the integrand
consists of only elementary functions, this can easily be done.

As an example consider the case $m_1=m_2=m_3=m_4=m_5=m_6=m$.
Now $|\M(x^2)|=6m^4x^2-2m^2x^4$ and 
\bea
|\M_J(x^2)|&=&3m^4J_1-x^2m^2(J_2+J_3)\nn\\
&=&{m\over(4\pi)^3}\left[\log{4\over 3}-\log{3m+x\over 2m+x}\right.\nn\\
&&
\left.
+{x\over 2m+x}\left({x\over 2m-x}\log{4m\over 2m+x}-\log{3m+x\over 3m}\right)
\right].
\eea 

Since $m_0$ can be chosen to be zero, the result is
\bea
\lefteqn{\In_A(m,m,m,m,m,m)=}\nn\\
&&
{1\over (4\pi)^3m^3}{1\over\sqrt{2}}
\int_0^1{dx\over\sqrt{3-x^2}}
\biggl(\log{3\over 4}+\log{3+x\over 2+x}\nn\\
&&-{x^2\over 4-x^2}\log{4\over 2+x}+{x\over 2+x}\log{3+x\over 3}
\biggr).
\eea

Numerical evaluation of this integral gives 
\beq
\In_A(m,m,m,m,m,m)\approx {0.0217376\over (4\pi)^3m^3}.
\eeq

\subsection{Master integral}

The integral $\In_a$ is called the master integral, since all the other two
loop two point integrals can be obtained from it by removing some of the
propagators from the integrand. Since $\In_a$ is the two point counterpart of 
$\In_A$ and it converges, it
can be calculated using Lemma \ref{rellemma}:
\bea
\lefteqn{\In_a(p;m_1,m_2,m_3,m_4,m_5)=}\nn\\
&&{2\pi \I\over p}
\left(
\In_A(m_3,m_1,m_2,\I p,m_5,m_4)-\In_A(m_3,m_1,m_2,-\I p,m_5,m_4)
\right).
\eea

In case of two point diagrams it is more convenient to use the following
notation
\bea
\label{Mpnotaatio}
\lefteqn{\M(p;m_1,m_2,m_3,m_4,m_5)\equiv\M(m_3,m_1,m_2,\I p,m_5,m_4)}\nn\\
&&=\pmatrix{2m_3^2&m_1^2+m_3^2-m_4^2&m_2^2+m_3^2-m_5^2\cr
            m_1^2+m_3^2-m_4^2&2m_1^2&m_1^2+m_2^2+p^2\cr
            m_2^2+m_3^2-m_5^2&m_1^2+m_2^2+p^2&2m_2^2}
\eea
and again, for simplicity
\beq
\M(p;m_3^2)\equiv\M(p;m_1,m_2,m_3,m_4,m_5).
\eeq
In the matrix $\M$ only the squares of masses are present and therefore
those are equal for both terms. One obtains
\bea
\lefteqn{\In_a(m_3^2) =
-{2\over\sqrt{|\M(p;m_3^2)|}}
\int_{m_0}^{m_3}dxx{1\over\sqrt{|\M(p;x^2)|}}}\nn\\
&&
\left(|\M_J(x,m_1,m_2,\I p,m_5,m_4)|-|\M_J(x,m_1,m_2,-\I p,m_5,m_4)|\right).
\label{Iaresult}
\eea

Now the matrices $\M_J$ differ only in the first row. Hence 
\beq
{2\pi \I\over p}(|\M_J(\ldots,\I p,\ldots)|-|\M_J(\ldots,-\I p,\ldots)|)
=|\M_H(\ldots,\I p,\ldots)|,
\eeq
where
\beq
H(\ldots,\I p,\ldots)={2\pi \I\over p}\left(
J(\ldots,\I p,\ldots)-J(\ldots,-\I p,\ldots)\right).
\eeq
Since vector $J$ consists of convergent vacuum integrals, one can use
Lemma \ref{rellemma} to notice that $H$ consists of corresponding
two point integrals:
\bea
\label{Hdef}
H_1(p;m_1,\ldots,m_5)&=&H_1^3-H_1^4+H_2^3-H_2^5\nn\\
H_2(p;m_1,\ldots,m_5)&=&H_2^1+H_3^1-H_3^4\nn\\
H_3(p;m_1,\ldots,m_5)&=&H_3^2-H_3^5+H_1^2,
\eea
where $H_i^j$ is, similarly to $J_i^j$, integral $\In_f$ or $\In_h$
which has been
constructed by removing the particle $j$ from integral $\In_a$ and taking 
square of 
the propagator
of particle $i$. The explicit expressions of these functions are given in
appendix \ref{appint}.

If one writes
\beq
\M_H(p;m_1,m_2,m_3,m_4,m_5)\equiv\M_H(m_3,m_1,m_2,\I p,m_5,m_4),
\eeq
Eq. (\ref{Iaresult}) can be written in the form
\bea
\lefteqn{\In_a(p;m_1,m_2,m_3,m_4,m_5)=}\nn\\
&&
-{2\over\sqrt{|\M(p;m_3^2)|}}
\int_{m_0}^{m_3}{|\M_H(p;m_1,m_2,x,m_4,m_5)|\over\sqrt{|\M(p;x^2)|}}xdx,
\label{IaresultOK}
\eea
where $m_0$ is again a root of the equation
\beq
|\M(p;m_0^2)|=0.
\eeq

As an example, consider now the case $m_1=\cdots=m_5=m$. Then the matrix 
$\M(p;x^2)=\M(p;m,m,x,m,m)$ 
is
\beq
\M(p;x^2)=\pmatrix{2x^2&x^2&x^2\cr x^2&2m^2&2m^2+p^2\cr x^2&2m^2+p^2&2m^2},
\eeq
and the functions $H_i$ are
\bea
H_1&=&{1\over(4\pi)^2mp^2(p^2+4m^2)}\nn\\
&&\biggl(
p\left(\arctan{p\over 2m}-\arctan{p\over 2m+x}\right)
-m\log{p^2+(2m+x)^2\over (2m+x)^2}\biggr),\nn\\
H_2&=&H_3={1\over (4\pi)^2 2mpx(2m+x)}
\biggl( \arctan{p\over 2m}-\arctan{p\over 2m+x}\biggr).
\eea

Substituting these to Eq. (\ref{IaresultOK}) gives the
result
\bea
\lefteqn{\In_a(p;m,m,m,m,m)=}\nn\\
&&{1\over (4\pi)^2 mp^2\sqrt{p^2+3m^2}}
\int_0^m{dx\over\sqrt{(p^2+4m^2)-x^2}}\nn\\
&&
\biggr[
{2p\over 2m+x}
  \left(\arctan{p\over 2m+x}-\arctan{p\over 2m}\right)
+\log{p^2+(2m+x)^2 \over (2m+x)^2}
\biggl].
\eea

\section{Effective potential and self energy in scalar theory
\label{scalarquant}}
\subsection{Lagrangian}
As an application of the integrals calculated in the previous section
scalar $\lambda\phi^4$ 
theory will now be discussed. This is the simplest possible 
nontrivial quantum field theory. However, it has also physical significance.
In theory of critical phenomena it is in the same universality class
as the Ising model.
The
Lagrangian of the theory is
\beq
{\cal L}=
{1\over 2}(\partial_i\phi)^2+{1\over 2}m_0^2\phi^2+{1\over 4}\lambda\phi^4.
\eeq

Suppose now that $m_0^2<0$. Then the minimum of the Lagrangian is not
anymore in the origin. Now make a shift $\phi\rightarrow \phi_0+\phi$
to get a new broken Lagrangian. The terms linear in fields are discarded,
since they cancel the tadpole terms in the true minimum when calculating
the self energy. When calculating the effective potential they are also
discarded, so that a presentation in terms of vacuum diagrams can be 
obtained \cite{ref:Jackiw}. The Lagrangian now reads
\beq
{\cal L}={1\over 2}m_0^2\phi_0^2+{1\over 4}\lambda\phi_0^4
+{1\over 2}(\partial_i\phi)^2+{1\over 2}m^2\phi^2
+\lambda\phi_0\phi^3+{1\over 4}\lambda\phi^4,
\eeq
where
\beq
m^2=m_0^2+3\lambda\phi_0^2.
\eeq

\begin{fmffile}{feynrules}

\begin{figure}
$$
\fmfframe(0,0)(0,-20){
\begin{fmfgraph*}(50,50)
\fmfleft{v1}
\fmfright{v2}
\fmf{plain_arrow,label=p}{v1,v2}
\end{fmfgraph*}}
={1\over p^2+m^2}
\hskip0.5cm
\fmfframe(0,0)(0,-20){
\begin{fmfgraph}(50,50)
\fmftop{v1}
\fmfbottom{v2,v3}
\fmf{plain}{v1,i1}
\fmf{plain}{v2,i1}
\fmf{plain}{v3,i1}
\end{fmfgraph}}
=\lambda\phi_0
\hskip0.5cm
\fmfframe(0,0)(0,-20){
\begin{fmfgraph}(50,50)
\fmftop{v0,v1}
\fmfbottom{v2,v3}
\fmf{plain}{v0,i1}
\fmf{plain}{v1,i1}
\fmf{plain}{v2,i1}
\fmf{plain}{v3,i1}
\end{fmfgraph}}
={1\over 4}\lambda
$$
\vskip1cm
\caption{
\label{scalarfeyn}
Feynman rules of the scalar theory}
\end{figure}

\end{fmffile}

The Feynman rules of the theory are
shown in Fig. \ref{scalarfeyn}.

\subsection{Self energy and renormalization}

\begin{table}
\centering{
\begin{tabular}{|r|l||r|l||r|l|}
\hline
$\alpha$&1&a&15552&A&1296\\
$\beta$&36&b&15552&B&1944\\
$\gamma$&6&c&864&C&216\\
&&d&432&D&216\\
&&e&288&E&24\\
&&f&864&F&72\\
&&g&192&&\\
&&h&432&&\\
\hline
\end{tabular}}
\caption{\label{setaulu}Contraction numbers of diagrams}
\end{table}

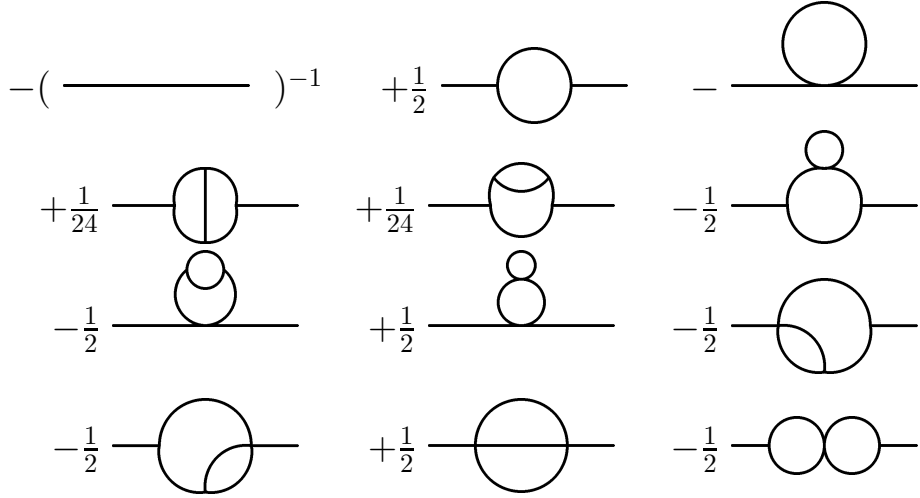
\begin{figure}[t]
\begin{fmffile}{skalself}
\large{$\Pi(k)=$}

\begin{tabular}{rrrr}
%\large{$\Pi=$}&
&\large{$-($}
\fmfframe(0,0)(0,-30){
\begin{fmfgraph}(70,70)
\fmfleft{v1}
\fmfright{v2}
\fmf{plain}{v1,i1}
\fmf{plain}{i1,v2}
\end{fmfgraph}
}\large{$)^{-1}$}
&
\large{$+{1\over 2}$}
\fmfframe(0,0)(0,-30){
\begin{fmfgraph}(70,70)
\fmfleft{v1}
\fmfright{v2}
\fmf{plain}{v1,i1}
\fmf{plain}{i2,v2}
\fmf{plain,left}{i1,i2}
\fmf{plain,right}{i1,i2}
\fmfforce{(.3w,.5h)}{i1}
\fmfforce{(.7w,.5h)}{i2}
\end{fmfgraph}}
&
\large{$-$}
\fmfframe(0,0)(0,-30){
\begin{fmfgraph}(70,70)
\fmfleft{v1}
\fmfright{v2}
\fmf{plain}{v1,i1}
\fmf{plain}{i1,v2}
\fmf{plain,left}{i1,i2}
\fmf{plain,left}{i2,i1}
\fmffreeze
\fmfforce{(.5w,.95h)}{i2}
\end{fmfgraph}
}
\\
&\large{$+{1\over 24}$}
\fmfframe(0,0)(0,-30){
\begin{fmfgraph}(70,70)
\fmfleft{v1}
\fmfright{v2}
\fmf{plain}{v1,i1}
\fmf{plain}{i3,v2}
\fmf{plain,left=0.5}{i1,i2}
\fmf{plain,left=0.5}{i2,i3}
\fmf{plain,right=0.5}{i1,i4}
\fmf{plain,right=0.5}{i4,i3}
\fmf{plain}{i2,i4}
\fmfforce{(.5w,.7h)}{i2}
\fmfforce{(.5w,.3h)}{i4}
\end{fmfgraph}
}
&
\large{$+{1\over 24}$}
\fmfframe(0,0)(0,-30){
\begin{fmfgraph}(70,70)
\fmfleft{v1}
\fmfright{v2}
\fmf{plain}{v1,i1}
\fmf{plain}{i4,v2}
\fmf{plain,left=0.2}{i1,i2}
\fmf{plain,left=0.2}{i3,i4}
\fmf{plain,right}{i1,i4}
\fmffreeze
\fmf{plain,left=0.5}{i2,i3}
\fmf{plain,right=0.5}{i2,i3}
\fmfforce{(.35w,.65h)}{i2}
\fmfforce{(.65w,.65h)}{i3}
\end{fmfgraph}
}
&
\large{$-{1\over 2}$}
\fmfframe(0,0)(0,-30){
\begin{fmfgraph}(70,70)
\fmfleft{v1}
\fmfright{v2}
\fmf{plain}{v1,i1}
\fmf{plain}{i3,v2}
\fmf{plain,left=0.5}{i1,i2}
\fmf{plain,left=0.5}{i2,i3}
\fmf{plain,right}{i1,i3}
\fmffreeze
\fmfforce{(.5w,.7h)}{i2}
\fmfforce{(.3w,.5h)}{i1}
\fmfforce{(.7w,.5h)}{i3}
\fmfforce{(.5w,.9h)}{i4}
\fmf{plain,right}{i2,i4,i2}
\end{fmfgraph}
}
\\
&
\large{$-{1\over 2}$}
\fmfframe(0,0)(0,-30){
\begin{fmfgraph}(70,70)
\fmfleft{v1}
\fmfright{v2}
\fmf{plain}{v1,i1}
\fmf{plain}{i1,v2}
\fmf{plain,left=0.7}{i1,i2}
\fmf{plain,left}{i2,i3}
\fmf{plain,right}{i2,i3}
\fmf{plain,left=0.7}{i3,i1}
\fmffreeze
\fmfforce{(.4w,.8h)}{i2}
\fmfforce{(.6w,.8h)}{i3}
\end{fmfgraph}
}&
\large{$+{1\over 2}$}
\fmfframe(0,0)(0,-30){
\begin{fmfgraph}(70,70)
\fmfleft{v1}
\fmfright{v2}
\fmf{plain}{v1,i1}
\fmf{plain}{i1,v2}
\fmf{plain,left}{i1,i2}
\fmf{plain,left}{i2,i3,i2}
\fmf{plain,left}{i2,i1}
\fmffreeze
\fmfforce{(.5w,.75h)}{i2}
\fmfforce{(.5w,.9h)}{i3}
\end{fmfgraph}
}&
\large{$-{1\over 2}$}
\fmfframe(0,0)(0,-30){
\begin{fmfgraph}(70,70)
\fmfleft{v1}
\fmfright{v2}
\fmf{plain}{v1,i1}
\fmf{plain}{i2,v2}
\fmf{plain,left}{i1,i2}
\fmf{plain,right=0.5}{i1,i3}
\fmf{plain,left=0.5}{i1,i3}
\fmf{plain,right=0.5}{i3,i2}
\fmffreeze
\fmfforce{(.5w,.25h)}{i3}
\fmfforce{(.25w,.5h)}{i1}
\fmfforce{(.75w,.5h)}{i2}
\end{fmfgraph}
}
\\
&
\large{$-{1\over 2}$}
\fmfframe(0,0)(0,-30){
\begin{fmfgraph}(70,70)
\fmfleft{v1}
\fmfright{v2}
\fmf{plain}{v1,i1}
\fmf{plain}{i2,v2}
\fmf{plain,left}{i1,i2}
\fmf{plain,right=0.5}{i1,i3}
\fmf{plain,left=0.5}{i3,i2}
\fmf{plain,right=0.5}{i3,i2}
\fmffreeze
\fmfforce{(.5w,.25h)}{i3}
\fmfforce{(.25w,.5h)}{i1}
\fmfforce{(.75w,.5h)}{i2}
\end{fmfgraph}
}
&
\large{$+{1\over 2}$}
\fmfframe(0,0)(0,-30){
\begin{fmfgraph}(70,70)
\fmfleft{v1}
\fmfright{v2}
\fmf{plain}{v1,i1}
\fmf{plain}{i2,v2}
\fmf{plain,left}{i1,i2}
\fmf{plain}{i1,i2}
\fmf{plain,right}{i1,i2}
\fmfforce{(.25w,.5h)}{i1}
\fmfforce{(.75w,.5h)}{i2}
\end{fmfgraph}
}
&
\large{$-{1\over 2}$}
\fmfframe(0,0)(0,-30){
\begin{fmfgraph}(70,70)
\fmfleft{v1}
\fmfright{v2}
\fmf{plain}{v1,i1}
\fmf{plain}{i3,v2}
\fmf{plain,left}{i1,i2}
\fmf{plain,right}{i1,i2}
\fmf{plain,left}{i2,i3}
\fmf{plain,right}{i2,i3}
\fmfforce{(.2w,.5h)}{i1}
\fmfforce{(.8w,.5h)}{i3}
\end{fmfgraph}
}
\end{tabular}
\end{fmffile}
\caption{
\label{sediag}
\label{seexp}Loop expansion of the scalar self energy}

\end{figure}

Let us start analyzing the system by calculating the self energy. In Fig. 
\ref{seexp}
it is expanded to two loop order. 
The number of different Wick contractions
corresponding to each diagram is given in table \ref{setaulu}.

In dimensional regularization only diagram g diverges. To remove this 
divergence a mass counterterm $\delta m^2$ must be introduced. Using the
given symmetry factors and integrals, the value of the divergent diagram is
\beq
{6\lambda^2\over (4\pi)^2}\left({1\over 4\ve}+{3\over 2}-
{3m\over p}\arctan{p\over 3m}+{1\over 2}\log{\bar\mu^2\over 9m^2+p^2}\right).
\eeq
The correct value of the mass counterterm is then
\beq
\delta m^2={\lambda^2\over (4\pi)^2}{3\over2\ve}.
\eeq
This gives rise to a running mass
\beq
m_0^2(\bar\mu)= {6\lambda^2\over (4\pi)^2}\log{\bar\mu\over \Lambda_m},
\eeq
where $\Lambda_m$ is a dimensional parameter such that $m_0^2(\Lambda_m)=0$.
Thus, the mass $m^2$ is a function of both the renormalization point and
the vacuum expectation value of the field:
\beq
m^2=m^2(\bar\mu,\phi_0).
\eeq
Since no other divergences are present, the coupling constant $\lambda$ does
not run.

The self energy can now be calculated by collecting all the relevant integrals
and the two loop part of the result is
\bea
\lefteqn{\Pi^{(2)}=}\nn\\
&&{1\over (4\pi)^2}
\left\{
\lambda^2
\left[
{9\over 2}-18{m\over p}\arctan{p\over 3m}+3\log{\bar\mu^2\over 9m^2+p^2}
\right]
\right.
\nn\\
&&
\left.
+\lambda^3\phi_0^2
\left[
{54\over p^2+4m^2}-{9\over m^2}-{54\over k^2}
\left(
\arctan{p\over 2m}
\right)
^2
\right.
\right.
\nn\\
&&
\left.
\left.
-{54\over pm}
\left[
2\log3\arctan{p\over 2m}
+\I
\left(
\Li2
\left(
-{\I p\over 3m}
\right)
+\Li2
\left(
-{2m-\I p\over m}
\right)
\right.
\right.
\right.
\right.
\nn\\
&&
\left.
\left.
\left.
\left.
-\Li2
\left(
{\I p\over 3m}
\right)
-\Li2
\left(
-{2m+\I p\over m}
\right)
\right)
\right]
\right]
\right.
\nn\\
&&
\left.
+\lambda^4\phi_0^4
\left[
{27\over m^3p^2(p^2+4m^2)}
\left(
4p(2p^2+11m^2)\arctan{p\over 3m}
\right.
\right.
\right.
\nn\\
&&
\left.
\left.
\left.
+(6\log 3-8)p(p^2+4m^2)\arctan{p\over 2m}
-6m(p^2+2m^2)\log
\left(1+{p^2\over 9m^2}\right)
\right.
\right.
\right.
\nn\\
&&
\left.
\left.
\left.
+3\I p(p^2+4m^2)
\left(
\Li2
\left(-{\I p\over 3m}\right)
-\Li2
\left({\I p\over 3m}\right)
\right.
\right.
\right.
\right.
\nn\\
&&
\left.
\left.
\left.
\left.
\Li2\left(-2+{\I p\over m}\right)-\Li2\left(-2-{\I p\over m}\right)
\right)
\right.
\right.
\right.
\nn\\
&&
\left.
\left.
\left.
+{648\over mp^2\sqrt{p^2+3m^2}}
\int_0^m{dx\over\sqrt{(p^2+4m^2)-x^2}}
\right.
\right.
\right.
\nn\\
&&
\left.
\left.
\left.
\left(
{2p\over 2m\!+\!x}
\left(\arctan{p\over 2m\!+\!x}-\arctan{p\over 2m}\right)
+\log{p^2+(2m\!+\!x)^2\over (2m\!+\!x)^2}
\right)
\right)
\right]
\right\}.
\eea

\subsection{Effective potential}
The calculation of the effective potential is very similar to that of 
the self energy. The diagrams to three loop order are given in Fig. 
\ref{epexp} and
the numbers of contractions in table \ref{setaulu}.

\begin{fmffile}{epot}

\begin{figure}
\begin{eqnarray*}
V(\phi_0) &=&
{1\over 2}m_0^2\phi_0^2+{1\over 4}\lambda\phi_0^4-{1\over 12\pi}m^3
+
\fmfframe(0,0)(0,-10){
\begin{fmfgraph}(20,30)
\fmftop{v1}
\fmfbottom{v2}
\fmf{plain,left}{i1,v1,i1}
\fmf{plain,left}{i1,v2,i1}
\end{fmfgraph}}
-{1\over 2}
\fmfframe(0,0)(0,-8){
\begin{fmfgraph}(20,20)
\fmfsurround{v1,v2}
\fmf{plain,right}{v1,v2}
\fmf{plain,left}{v1,v2}
\fmf{plain}{v1,v2}
\end{fmfgraph}}
\\
&&
-{1\over 2}
\fmfframe(0,10)(0,-12){
\begin{fmfgraph}(30,30)
\fmfsurround{v1,v2}
\fmf{plain,right}{v1,v2}
\fmf{plain,left}{v1,v2}
\fmf{plain,right=0.4}{v1,v2}
\fmf{plain,left=0.4}{v1,v2}
\end{fmfgraph}}
-{1\over 2}
\fmfframe(0,10)(0,-12){
\begin{fmfgraph}(35,30)
\fmfsurround{v1,v2}
\fmf{plain,right}{v1,i1}
\fmf{plain,left}{v1,i1}
\fmf{plain,right}{i1,i2}
\fmf{plain,left}{i1,i2}
\fmf{plain,right}{i2,v2}
\fmf{plain,left}{i2,v2}
\end{fmfgraph}}
+{1\over 2}
\fmfframe(5,10)(0,-12){
\begin{fmfgraph}(30,30)
\fmfsurround{v1,v2,v3}
\fmf{plain,right}{v1,i1}
\fmf{plain,left}{v1,i1}
\fmf{plain,right=0.5}{i1,v2}
\fmf{plain,right=0.5}{v3,i1}
\fmf{plain,right}{v2,v3}
\fmf{plain}{v2,v3}
\end{fmfgraph}}
+{1\over 2}
\fmfframe(0,10)(0,-12){
\begin{fmfgraph}(30,30)
\fmfsurround{v3,v0,v00,v1,v2,vo,voo}
\fmf{plain,left=0.7}{v1,v3}
\fmf{plain}{v2,v3}
\fmf{plain}{v1,v3}
\fmf{plain,right=0.7}{v2,v3}
\fmf{plain,right=0.3}{v1,v2}
\end{fmfgraph}}\\
&&
-{1\over 24}
\fmfframe(0,10)(0,-12){
\begin{fmfgraph}(30,30)
\fmfsurround{v0,v1,v2,v00,v3,v4}
\fmf{plain,left=0.5}{v1,v4}
\fmf{plain,left=0.3}{v2,v3}
\fmf{plain,right=0.3}{v1,v4}
\fmf{plain,right=0.5}{v2,v3}
\fmf{plain,right=0.2}{v1,v2}
\fmf{plain,right=0.2}{v3,v4}
\end{fmfgraph}}
-{1\over 24}
\fmfframe(0,10)(0,-12){
\begin{fmfgraph}(30,30)
\fmfsurround{v1,v2,v3}
\fmf{plain}{v1,i1}
\fmf{plain}{v2,i1}
\fmf{plain}{v3,i1}
\fmf{plain,right=0.5}{v1,v2}
\fmf{plain,right=0.5}{v2,v3}
\fmf{plain,right=0.5}{v3,v1}
\end{fmfgraph}}
\end{eqnarray*}

\caption{\label{epexp}Loop expansion of the effective potential}

\

\end{figure}

\end{fmffile}

The result can be written down at once
\bea
V(\phi_0)&=&
{1\over 2}m_0^2\phi_0^2+{1\over 4}\lambda\phi_0^4-{1\over 12\pi}m^3\nn\\
&&+{1\over (4\pi)^2}
\left[
{3\over 4}\lambda m^2
-3\lambda^2\phi_0^2
\left(\log{\bar\mu\over 3m}+{1\over 2}\right)
\right]
\nn\\
&&+
{1\over (4\pi)^3}
\left\{
m\lambda^2
\left(3\log{\bar\mu\over 4m}+{27\over 8}\right)
\right.
\nn\\
&&
\left.
+{\lambda^3\phi_0^2\over m}
\left[
-{9\over 2}+{9\over 4}\pi^2
-{27\over 2}
\left(\log{4\over3}\right)
^2
-27\Li2
\left({1\over 4}\right)
\right]
\right.
\nn\\
&&
\left.
+{\lambda^4\phi_0^4\over m^3}
\left[
-{27\over 8}\pi^2
+{81\over 4}
\left(\log{4\over3}\right)
^2
+54\log{4\over 3}+{81\over 2}\Li2
\left({1\over 4}\right)
\right.
\right.
\nn\\
&&
\left.
\left.
-54
{1\over\sqrt{2}}
\int_0^1{dx\over\sqrt{3-x^2}}
\left(\log{3\over 4}+\log{3+x\over 2+x}
\right.
\right.
\right.
\nn\\
&&
\left.
\left.
\left.
-{x^2\over 4-x^2}\log{4\over 2+x}+{x\over 2+x}\log{3+x\over 3}
\right)
\right]
\right\}.
\eea

This result agrees perfectly with that evaluated numerically in 
\cite{ref:mersu}.

\section{Conclusions}
In this paper, the integrals necessary for the self energy to two loop level
in a three dimensional scalar field theory 
have been evaluated explicitly as well as the ones necessary for the
effective potential to three loop level. In almost every case the result can
be expressed with elementary functions and dilogarithms. 

A large part of the paper has been devoted to the evaluation of the
two most difficult
integrals, the master integral and the Mercedes integral.
Those are expressed in
terms of a one dimensional integral representation with an integrand consisting
only of elementary functions. This form makes numerical evaluation easy.

The results have been applied to $\lambda\phi^4$ 
scalar theory. 
It will be very interesting to extend them to gauge theories with
scalars, like the U(1)+Higgs or SU(2)+Higgs models. This will lead to
a large number of new diagrams.
However,
all the two loop integrals contributing to the
self energy in a gauge field theory can be decomposed to a sum of
scalar integrals. This is a laborious task for which a computer algebra
system is needed. 
The results will help to deepen our understanding of the phase transitions
in gauge theories, for example the electroweak phase transition of the
early universe.

\section*{Acknowledgements}
The author wishes to thank K.~Kajantie for useful discussions.
Financial support by the Academy of Finland is also gratefully acknowledged.

\appendix
\section{Integrals}
\label{appint}
This is a complete list of the integrals corresponding to the diagrams
of Figs. \ref{topo1l}, \ref{topose} and \ref{topoep}.
\bea
\alpha) &=& -{m\over 4\pi}
\left[1+\ve\left(2+2\log{\bar\mu\over 2m}\right)\right].\\
\beta)  &=& {1\over 4\pi p}\arctan{p\over m_1+m_2}.\\
\gamma) &=& {1\over (4\pi)^2}\left({1\over 4\ve}+{1\over 2}+
\log{\bar\mu\over m_1+m_2+m_3}\right).\\
a)&=& 
-{2\over\sqrt{|\M(p;m_1,m_2,m_3,m_4,m_5)|}}\nn\\
&&
\int_{m_0}^{m_3}{|\M_H(p;m_1,m_2,x,m_4,m_5)|
\over\sqrt{|\M(p;m_1,m_2,x,m_4,m_5)|}}xdx,
\eea
where
\bea
\lefteqn{\M(p;m_1,m_2,m_3,m_4,m_5)=}\nn\\
&&\pmatrix{2m_3^2&m_1^2+m_3^2-m_4^2&m_2^2+m_3^2-m_5^2\cr
            m_1^2+m_3^2-m_4^2&2m_1^2&m_1^2+m_2^2+p^2\cr
            m_2^2+m_3^2-m_5^2&m_1^2+m_2^2+p^2&2m_2^2},
\eea
and
\bea
\lefteqn{\M_H(p;m_1,m_2,m_3,m_4,m_5)=}\nn\\
&&\pmatrix{H_1&H_2&H_3\cr
            m_1^2+m_3^2-m_4^2&2m_1^2&m_1^2+m_2^2+p^2\cr
            m_2^2+m_3^2-m_5^2&m_1^2+m_2^2+p^2&2m_2^2}.
\eea
The functions $H_i(p;m_1,m_2,m_3,m_4,m_5)$ are
\bea
H_1(p;m_1,\ldots,m_5)&=&H_1^3-H_1^4+H_2^3-H_2^5\nn\\
H_2(p;m_1,\ldots,m_5)&=&H_2^1+H_3^1-H_3^4\nn\\
H_3(p;m_1,\ldots,m_5)&=&H_3^2-H_3^5+H_1^2,
\eea
where
\bea
H_1^3&=&H_X(m_1,m_2,m_4,m_5)\nn\\
H_1^4&=&H_Y(m_1,m_3,m_5,m_2)\nn\\
H_2^3&=&H_X(m_2,m_1,m_5,m_4)\nn\\
H_2^5&=&H_Y(m_2,m_3,m_4,m_1)\nn\\
H_2^1&=&H_Z(m_4,m_2,m_3,m_5)\nn\\
H_3^1&=&H_Z(m_4,m_3,m_2,m_5)\nn\\
H_3^4&=&H_Z(m_1,m_3,m_5,m_2)\nn\\
H_3^2&=&H_Z(m_5,m_3,m_1,m_4)\nn\\
H_3^5&=&H_Z(m_2,m_3,m_4,m_1)\nn\\
H_1^2&=&H_Z(m_5,m_1,m_3,m_4),
\eea
and
\bea
\lefteqn{H_X(m_1,m_2,m_3,m_4)=}\nn\\
\phantom{=}&&
{1\over (4\pi)^2}{1\over 2m_1p\left((m_1+m_2)^2+p^2\right)}
\arctan{p\over m_3+m_4}\nn\\
\lefteqn{H_Y(m_1,m_2,m_3,m_4)=}\nn\\
&&
{1\over (4\pi)^2}{1\over 2m_1p\left((p^2+m_1^2+m_4^2)^2-4m_1^2m_2^2
\right)}\nn\\
&&
\left((p^2+m_4^2-m_1^2)\arctan{p\over m_1+m_2+m_3}\right.\nn\\
&&\left.
       +m_1p\log{p^2+(m_1+m_2+m_3)^2\over (m_2+m_3+m_4)^2}
       \right)\nn\\
\lefteqn{H_Z(m_1,m_2,m_3,m_4)=}\nn\\
&&
{1\over (4\pi)^2}{1\over 2m_2p(m_4^2-(m_2+m_3)^2)}\nn\\
&&\left(\arctan{p\over m_1+m_2+m_3}-\arctan{p\over m_1+m_4}\right).
\eea

\bea
b)&=& 
{1\over (4\pi)^2 4 p m_1 m_4 (m_4^2-m_1^2)}
\left\{
 2m_1\log{m_2+m_3+m_4 \over m_2+m_3-m_4}\arctan{p\over m_4+m_5}
\right.
\nn\\
&&
\left.
+2m_4\log{m_1+m_2+m_3 \over m_2+m_3-m_1}\arctan{p\over m_1+m_5}
\right.
\nn\\
&&
\left.
+\I
m_4
\left[
\Li2\left(-{m_5+m_1-\I p\over m_2+m_3-m_1}\right)+
\Li2\left(-{m_5+m_1+\I p\over m_2+m_3-m_1}\right)
\right.
\right.
\nn\\
&&
\left.
\left.
-
\Li2\left(-{m_5-m_1-\I p\over m_1+m_2+m_3}\right)-
\Li2\left(-{m_5-m_1+\I p\over m_1+m_2+m_3}\right)
\right]
\right.
\nn\\
&&
\left.
+\I m_1
\left[
\Li2\left(-{m_5-m_4+\I p\over m_2+m_3+m_4}\right)+
\Li2\left(-{m_5-m_4-\I p\over m_2+m_3+m_4}\right)
\right.
\right.
\nn\\
&&
\left.
\left.
-
\Li2\left(-{m_5+m_4+\I p\over m_2+m_3-m_4}\right)-
\Li2\left(-{m_5+m_4-\I p\over m_2+m_3-m_4}\right)
\right]
\right\}.\\
c)&=&{m_2\over (4\pi)^2p(m_1^2-m_3^2)}\left(
\arctan{p\over m_1+m_4}-\arctan{p\over m_3+m_4}\right).\\
d)&=&{1\over (4\pi)^2(m_4^2-m_1^2)}\log{m_2+m_3+m_4\over m_1+m_2+m_3}.\\
e)&=&-{1\over (4\pi)^2}{m_2\over m_1+m_3}.\\
f)&=&{1\over (4\pi)^2 4 p m_4}
\left\{
2\log{m_2+m_3+m_4 \over m_2+m_3-m_4}\arctan{p\over m_1+m_4}
\right.
\nn\\
&&
\left.
+\I
\left[
\Li2\left(-{m_1+m_4-\I p\over m_2+m_3-m_4}\right)-
\Li2\left(-{m_1-m_4-\I p\over m_2+m_3+m_4}\right)
\right.
\right.
\nn\\
&&
\left.
\left.
+
\Li2\left(-{m_1-m_4+\I p\over m_2+m_3+m_4}\right)-
\Li2\left(-{m_1+m_4+\I p\over m_2+m_3-m_4}\right)
\right]
\right\}.\\
g)&=&
{1\over(4\pi)^2}
\left(
{1\over 4\ve}+{3\over 2}-{m_1+m_2+m_3\over p}
\arctan{p\over m_1+m_2+m_3}
\right.
\nn\\
&&
\left.
+{1\over 2}\log{\bar\mu^2\over(m_1+m_2+m_3)^2+p^2}
\right)
.\\
h)&=&{1\over(4\pi)^2p^2}\arctan{p\over m_1+m_2}\arctan{p\over m_3+m_4}.\\
A)&=&-{2\over\sqrt{|\M(m_1,m_2,m_3,m_4,m_5,m_6)|}}\nn\\
&&
\int_{m_0}^{m_1}{|\M_J(x,m_2,m_3,m_4,m_5,m_6)|
\over\sqrt{|\M(x,m_2,m_3,m_4,m_5,m_6)|}}xdx,
\eea
where the matrices $\M$ and $\M_J$ are
\bea
\lefteqn{\M(m_1,m_2,m_3,m_4,m_5,m_6)=}\nn\\
&&\pmatrix{2m_1^2 & m_1^2+m_2^2-m_6^2 & m_1^2+m_3^2-m_5^2\cr
                m_1^2+m_2^2-m_6^2 & 2m_2^2 & m_2^2+m_3^2-m_4^2\cr
                m_1^2+m_3^2-m_5^2 & m_2^2+m_3^2-m_4^2 & 2m_3^2},
\eea
and
\bea
\lefteqn{\M_J(m_1,m_2,m_3,m_4,m_5,m_6)=}\nn\\
&&\pmatrix{J_1 & J_2 & J_3\cr
                m_1^2+m_2^2-m_6^2 & 2m_2^2 & m_2^2+m_3^2-m_4^2\cr
                m_1^2+m_3^2-m_5^2 & m_2^2+m_3^2-m_4^2 & 2m_3^2}.
\eea
The functions $J_i(m_1,m_2,m_3,m_4,m_5,m_6)$ are 
\bea
J_1&=&J_2^1-J_2^6+J_3^1-J_3^5\nn\\
J_2&=&J_3^2-J_3^4+J_1^2-J_1^6\nn\\
J_3&=&J_1^3-J_1^5+J_2^3-J_2^4,
\eea
where
\bea
J_2^1&=&\tilde\In_C(m_2,m_3,m_5,m_6,m_4)\nn\\
J_2^6&=&\tilde\In_C(m_2,m_4,m_1,m_5,m_3)\nn\\
J_3^1&=&\tilde\In_C(m_3,m_2,m_5,m_6,m_4)\nn\\
J_3^5&=&\tilde\In_C(m_3,m_4,m_1,m_6,m_2)\nn\\
J_3^2&=&\tilde\In_C(m_3,m_4,m_1,m_6,m_5)\nn\\
J_3^4&=&\tilde\In_C(m_3,m_5,m_2,m_6,m_1)\nn\\
J_1^2&=&\tilde\In_C(m_1,m_6,m_3,m_4,m_5)\nn\\
J_1^6&=&\tilde\In_C(m_1,m_5,m_2,m_4,m_3)\nn\\
J_1^3&=&\tilde\In_C(m_1,m_2,m_4,m_5,m_6)\nn\\
J_1^5&=&\tilde\In_C(m_1,m_6,m_3,m_4,m_2)\nn\\
J_2^3&=&\tilde\In_C(m_2,m_1,m_4,m_5,m_6)\nn\\
J_2^4&=&\tilde\In_C(m_2,m_6,m_3,m_4,m_1),
\eea
and $\tilde\In_C$ is the derivative of $\In_C$:
\bea
\lefteqn{
\tilde\In_C(m_1,m_2,m_3,m_4,m_5)=}\nn\\
&&
{1\over(4\pi)^32m_1
((m_1+m_2)^2-m_5^2)}
\log{m_1+m_2+m_3+m_4\over m_3+m_4+m_5}.
\eea

\bea
B)&=&{1\over (4\pi)^3 4 m_5m_6(m_6^2-m_5^2)}
\nn\\
&&
\left\{
m_6
\left[
\log{m_3+m_4+m_5\over m_3+m_4-m_5}
\log{(m_3+m_4)^2-m_5^2\over (m_1+m_2+m_5)^2}
\right.
\right.
\nn\\
&&
\left.
\left.
\!+\!2\Li2\left(-{m_1+m_2-m_5\over m_3+m_4+m_5}\right)
-2\Li2\left(-{m_1+m_2+m_5\over m_3+m_4-m_5}\right)
\right]
\right.
\nn\\
&&
\left.
-m_5
\left[
\log{m_3+m_4+m_6\over m_3+m_4-m_6}
\log{(m_3+m_4)^2-m_6^2\over (m_1+m_2+m_6)^2}
\right.
\right.
\nn\\
&&
\left.
\left.
+\!2\Li2\left(-{m_1+m_2-m_6\over m_3+m_4+m_6}\right)
-2\Li2\left(-{m_1+m_2+m_6\over m_3+m_4-m_6}\right)
\right]
\right\}
.\\
C)&=&
{1\over (4\pi)^3 2 m_5}
\left[\log{m_3+m_4+m_5\over m_3+m_4-m_5}
\log{\left(m_3+m_4)^2-m_5^2\right)^{1\over 2}\over m_1+m_2+m_5}
\right.\nn\\
&&
\left.
+{\rm Li_2}\left(-{m_1+m_2-m_5\over m_3+m_4+m_5}\right)
-{\rm Li_2}\left(-{m_1+m_2+m_5\over m_3+m_4-m_5}\right)
\right].\\
D)&=&
{m_5\over (4\pi)^3(m_3^2-m_4^2)}\log{m_1+m_2+m_4\over m_1+m_2+m_3}.\\
E)&=&
{1\over (4\pi)^3}\sum_{i=1}^4m_i
\left(-{1\over 4\ve}+2+{1\over 2}\log{2m_i\over \bar\mu}
+\log{\sum_jm_j\over\bar\mu}\right).
\\
F)&=&
{1\over (4\pi)^3}{m_1m_4\over m_2+m_3}.
\eea

\bibliographystyle{plain}
\bibliography{gradu}

\begin{thebibliography}{10}

%\bibitem{ref:ZJ}
%J.~Zinn-Justin,
%\newblock Quantum Field Theory and Critical Phenomena.
%\newblock Oxford Univ. Press, 1989.

%\bibitem{ref:Parisi}
%G.~Parisi,
%\newblock {Statistical Field Theory}.
%\newblock Addison-Wesley, 1988.

%\bibitem{ref:kapusta}
%J.I. Kapusta,
%\newblock {Finite-temperature field theory}.
%\newblock Cambridge Univ. Press, 1993.

\bibitem{ref:Ginsparg}
P.~Ginsparg,
\newblock Nucl. Phys. {\bf B 170} (1980) 388.

\bibitem{ref:JacTem}
R.~Jackiw and S.~Templeton,
\newblock Phys. Rev. {\bf D 23} (1981) 2291.

\bibitem{ref:Appelquist}
T.~Appelquist and R.~Pisarski,
\newblock Phys. Rev. {\bf D 23} (1981) 2305.

\bibitem{ref:Nadkarni}
S.~Nadkarni,
\newblock Phys. Rev. {\bf D 27} (1983) 917.
 
\bibitem{ref:Landsman}
N.P.~Landsman,
\newblock Nucl. Phys. {\bf B 322} (1989) 498.

\bibitem{ref:dimred}
K.~Kajantie, M.~Laine, K.~Rummukainen, and M.~Shaposhnikov,
\newblock  Nucl. Phys. {\bf B 458} (1996) 90.

\bibitem{ref:electroweak}
K.~Farakos, K.~Kajantie, K.~Rummukainen, and M.~Shaposhnikov,
\newblock { Nucl. Phys. {\bf B 425} (1994) 67}.

\bibitem{ref:lattice}
K.~Kajantie, M.~Laine, K.~Rummukainen, and M.~Shaposhnikov,
\newblock { hep-lat/9510020}.

\bibitem{ref:uusi}
K.~Kajantie, M.~Laine, K.~Rummukainen, and M.~Shaposhnikov,
\newblock { hep-ph/9605288}.


\bibitem{ref:series}
A.I.~Davydychev and J.B.~Tausk,
\newblock { Nucl. Phys. {\bf B 397} (1993) 123}.

\bibitem{ref:series2}
D.J.~Broadhurst, J.~Fleischer, and O.V.~Tarasov,
\newblock { Z. Phys. {\bf C 60} (1993) 287}.

\bibitem{ref:Mellin}
{\'E.\'E.}~Boos and A.I.~Davydychev,
\newblock { Theor. Math. Phys. {\bf 89} (1992) 1052}.

\bibitem{ref:Gegenbauer}
P.N.~Maher, L.~Durand, and K.~Riesselmann,
\newblock { Phys. Rev. {\bf D 48} (1993) 1061}.

\bibitem{ref:Scharf}
R.~Scharf and J.B.~Tausk,
\newblock{ Nucl. Phys. {\bf B 412} (1994) 523.}

\bibitem{ref:Bauberger}
S.~Bauberger and M.~B{\"o}hm,
\newblock { Nucl. Phys. {\bf B 445} (1995) 25}.

\bibitem{ref:Kreimer}
D.~Kreimer,
\newblock { Phys. Lett. {\bf B 273} (1991) 277}.

\bibitem{ref:Ghinculov}
A.~Ghinculov and J.J. van~der Bij,
\newblock { Nucl. Phys. {\bf B 436} (1995) 30}.

\bibitem{ref:Avdeev}
L.V.~Avdeev,
\newblock { hep-ph/9512442}.

\bibitem{ref:Broadhurst}
D.J.~Broadhurst,
\newblock { Z. Phys. {\bf C 54} (1992) 599}.


\bibitem{ref:Weiglein}
G.~Weiglein, R.~Scharf, and M.~B{\"o}hm,
\newblock { Nucl. Phys. {\bf B 416} (1994) 606}.

\bibitem{ref:Jackiw}
R.~Jackiw,
\newblock { Phys. Rev. {\bf D 9} (1974) 1686}.

%\bibitem{ref:dimreg}
%G.~'t~Hooft and M.~Veltman,
%\newblock { Nucl. Phys. {\bf B 44} (1972) 189}.

\bibitem{ref:arnoldesp}
P.~Arnold and O.~Espinosa,
\newblock{ Phys. Rev. {\bf D 47} (1993) 3546.}

\bibitem{ref:fourier}
E.~Braaten and A.~Nieto,
\newblock { Phys. Rev. {\bf D 51} (1995) 6990}.

\bibitem{ref:Kotikov}
A.V.~Kotikov,
\newblock { Phys. Lett. {\bf B 254} (1991) 158},
\newblock { Mod. Phys. Lett. {\bf A 6} (1991) 677}.

\bibitem{ref:Zhai}
C.~Zhai and B.~Kastening,
\newblock { hep-ph/9507380}.

\bibitem{ref:mersu}
C.~Bagnuls, C.~Bervillier, D.I.~Meiron, and B.G.~Nickel,
\newblock { Phys. Rev. {\bf B 35} (1987) 3585}.

%\bibitem{ref:PVelt}
%M.~Veltman G.~Passarino,
%\newblock { Nucl. Phys. {\bf B 160} (1979) 151}.

\end{thebibliography}
\end{document}

\begin{frontmatter}
\title
{Erratum to ``Feynman diagrams to three loops in three-dimensional
field theory'' \\ {[Nucl. Phys. B 480 (1997) 729]}\thanksref{pii}}
\thanks[pii]{PII of the original article: S0550-3213(96)00474-9}
\author[email]{Arttu K. Rajantie}
\thanks[email]{Present address:\\
DAMTP, CMS, University of Cambridge,\\
Wilberforce Road, Cambridge CB3 0WA, United Kingdom\\
E-mail: a.k.rajantie@damtp.cam.ac.uk}
\address{Department of Physics, P.O. Box 9, FIN-00014 University of Helsinki, 
Finland} 
\maketitle
\end{frontmatter}
The sign of the constant term $-2$ was incorrect in 
Eqs.~(21) and (A.26):
\begin{equation}
{1\over (4\pi)^3}\sum_{i=1}^4m_i
\left(-{1\over 4\ve}-2+{1\over 2}\log{2m_i\over \bar\mu}
+\log{\sum_jm_j\over\bar\mu}\right).
\end{equation}

One $m_2$ should be $m_4$ in Eq.~(A.9):
\bea
H_Y(m_1,m_2,m_3,m_4)&=&
{1\over (4\pi)^2}{1\over 2m_1p\left((p^2+m_1^2+m_4^2)^2-4m_1^2m_4^2
\right)}\nn\\
&&
\times\left((p^2+m_4^2-m_1^2)\arctan{p\over m_1+m_2+m_3}\right.\nn\\
&&\left.
       +m_1p\log{p^2+(m_1+m_2+m_3)^2\over (m_2+m_3+m_4)^2}
       \right).
\eea

Five signs were incorrect in Eq.~(A.10):
\bea
(b)&=& 
{1\over (4\pi)^2 4 p m_1 m_4 (m_4^2-m_1^2)}
\left\{
2m_4\log{m_1+m_2+m_3 \over m_2+m_3-m_1}\arctan{p\over m_1+m_5}
\right.
\nn\\
&&
\left.
-2m_1\log{m_2+m_3+m_4 \over m_2+m_3-m_4}\arctan{p\over m_4+m_5}
\right.
\nn\\
&&
\left.
+\I
m_4
\left[
\Li2\left(-{m_5+m_1-\I p\over m_2+m_3-m_1}\right)-
\Li2\left(-{m_5+m_1+\I p\over m_2+m_3-m_1}\right)
\right.
\right.
\nn\\
&&
\left.
\left.
-
\Li2\left(-{m_5-m_1-\I p\over m_1+m_2+m_3}\right)+
\Li2\left(-{m_5-m_1+\I p\over m_1+m_2+m_3}\right)
\right]
\right.
\nn\\
&&
\left.
-\I m_1
\left[
\Li2\left(-{m_5+m_4-\I p\over m_2+m_3-m_4}\right)
-
\Li2\left(-{m_5+m_4+\I p\over m_2+m_3-m_4}\right)
\right.
\right.
\nn\\
&&
\left.
\left.
-
\Li2\left(-{m_5-m_4-\I p\over m_2+m_3+m_4}\right)
+
\Li2\left(-{m_5-m_4+\I p\over m_2+m_3+m_4}\right)
\right]
\right\}.
\eea

In three of the quantities $J^i_j$ listed in Eq.~(A.21),
one or two mass indices were incorrect:
\bea
J_3^2&=&\tilde\In_C(m_3,m_1,m_4,m_6,m_5),\nn\\
J_1^2&=&\tilde\In_C(m_1,m_3,m_4,m_6,m_5),\nn\\
J_2^4&=&\tilde\In_C(m_2,m_6,m_3,m_5,m_1).
\eea

In Eq.~(A.23) the $\pi^2/12$ terms were missing:
\bea
(B)&=&
{1\over (4\pi)^3 (m_6^2-m_5^2)}\nn\\
&&
\left\{
{1\over m_5}\left[\frac{\pi^2}{12}+
   {1\over 4}\left(\log {m_5+m_1+m_2 \over m_5+m_3+m_4}\right)^2
\right.\right.\nn\\
&&\left.\left.
  +{1\over 2}\Li2
    \left({m_5-m_1-m_2\over m_5+m_3+m_4}\right)
  +{1\over 2}\Li2
    \left({m_5-m_3-m_4\over m_5+m_1+m_2}\right)\right]\right.\nn\\
&&\left.
-{1\over m_6}\left[\frac{\pi^2}{12}+
   {1\over 4}\left(\log {m_6+m_1+m_2 \over m_6+m_3+m_4}\right)^2
\right.\right.\nn\\
&&\left.\left.
  +{1\over 2}\Li2
    \left({m_6-m_1-m_2\over m_6+m_3+m_4}\right)
  +{1\over 2}\Li2
    \left({m_6-m_3-m_4\over m_6+m_1+m_2}\right)\right]\right\}.
\eea

On p.~740 the sign of the integration constant $C$ should be the opposite.
On the fourth line of p.~742 there should be $\sqrt{|{\mathcal M}(x^2)|}$
and on p.~743, below Eq.~(59), $|{\mathcal M}(p;m_0^2)|=0$.

These errors change neither 
the results in Eqs.~(77) and (78) nor any of the conclusions.

I would like to thank F.~Eberlein, S.~Larin and M.~Stroesser for pointing
out these mistakes for me.

Reprinted from Nuclear Physics B 513, Arttu K.~Rajantie, 
{\em Erratum to ``Feynman diagrams to three loops in three-dimensional
field theory''}, pp.~761--762, Copyright (1998), with permission from
Elsevier Science.